\documentclass[twoside]{article}
\usepackage[a4paper]{geometry}
\usepackage[latin1]{inputenc} % ou \usepackage[utf8]{inputenc}
\usepackage[OT1]{fontenc} % ou \usepackage[OT1]{fontenc}
\usepackage{RR}
\usepackage{hyperref}
\usepackage{makecell}
\newcommand{\yes}{\cmark}

\newcommand{\cmplt}{\ensuremath{\CIRCLE}}
\newcommand{\prtl}{\ensuremath{\LEFTcircle}}
\newcommand{\absn}{\ensuremath{\Circle}}

\newcommand{\nlpp}{natural language privacy policies\xspace}
\newcommand{\Nlpp}{Natural language privacy policies\xspace}
\newcommand{\gpp}{graphical privacy policies\xspace}
\newcommand{\Gpp}{Graphical privacy policies\xspace}
\newcommand{\mrpp}{machine-readable privacy policies\xspace}
\newcommand{\Mrpp}{Machine-readable privacy policies\xspace}

\newcommand{\eg}{\textit{e.g.}\xspace}
\newcommand{\ie}{\textit{i.e.}\xspace}
\newcommand{\etal}{\textit{et al.}\xspace}
\newcommand{\etc}{\textit{etc.}\xspace}

\RRNo{9287}
\RRdate{September 2020}
\RRauthor{% les auteurs
  Victor Morel
  \and
  Ra\'ul Pardo
}
\authorhead{V. Morel \& R. Pardo}
\RRtitle{SoK : Trois facettes des politiques de protection de vie priv\'ee}
\RRetitle{SoK: Three Facets of Privacy Policies}
\titlehead{SoK: Three Facets of Privacy Policies}
\RRresume{
  Les \textit{politiques de protection de vie privée} sont le
  principal moyen d'obtenir de l'information liée à la collecte et au
  traitement de données à caractère personnel.
  Ces politiques \'etaient originellement pr\'esent\'ees comme des
  documents textuels.
  Cependant, la non-convenance de ce format aux besoins de la
  soci\'et\'e actuelle a donn\'e lieu à d'autres moyens d'expression.
  Dans ce rapport, nous \'etudions de mani\`ere syst\'ematique les
  diff\'erents moyens d'expression des politiques de protection de vie
  priv\'ee.
  Ce faisant, nous identifions trois cat\'egories principales que nous
  nommons \textit{dimensions}, \ie, les politiques en langage naturel,
  la repr\'esentation graphique des politiques, et les politiques
  lisibles par les machines.
  Chacune de ses dimensions se concentre sur les besoins sp\'ecifiques
  de la communaut\'e dont elle est issue, \ie, respectivement les
  juristes, les organisations et les d\'efenseurs de la vie priv\'ee,
  et les universitaires.
  Nous analysons ensuite les avantages et les limites de chaque
  dimension, et nous expliquons en quoi les solutions bas\'ees sur une
  seule dimension ne couvrent pas les besoins des autres
  communaut\'es.
  Enfin, nous proposons une nouvelle approche pour exprimer les
  politiques de protection de vie priv\'ee qui r\'eunit les avantages
  de chaque dimension, dans le but de surmonter leurs limites.
}

\RRabstract{
  \emph{Privacy policies} are the main way to obtain information
  related to personal data collection and processing.
  Originally, privacy policies were presented as textual documents.
  However, the unsuitability of this format for the needs of today's
  society gave birth to other means of expression.
  In this paper, we systematically study the different means of
  expression of privacy policies.
  In doing so, we have explored the three main categories, which we
  call \emph{facets}, \ie, natural language, graphical and
  machine-readable privacy policies.
  Each of these facets focuses on the particular needs of the
  communities they come from, \ie, law experts, organizations and
  privacy advocates, and academics, respectively.
  We then analyze the benefits and limitations of each facet,
  and explain why solutions based on a single facet do not cover
  the needs of other communities.
  Finally, we set guidelines and discuss challenges of an approach to
  expressing privacy policies which brings together the benefits of
  each facet as an attempt to overcome their limitations.
}
\RRmotcle{politiques de vie priv\'ee, conformit\'e l\'egale, utilisabilit\'e, mise en application}
\RRkeyword{privacy policies, legal compliance, usability, enforcement}
\RRprojets{Privatics}
\URRhoneAlpes % pour ceux qui sont dans les montagnes
\RCGrenoble % Grenoble - Rh\^one-Alpes
\PassOptionsToPackage{dvipsnames}{xcolor} %% To have more colours
\usepackage[numbers,square]{natbib}
\usepackage{listings}
\usepackage{graphicx}
\usepackage[english]{babel}
\usepackage{bibgerm}
\usepackage{nameref}
\usepackage{algorithm}
\usepackage{amsmath}
\usepackage{amssymb}
\usepackage{algorithmic}
\usepackage{mathtools}
\usepackage{amsthm}
\usepackage{subcaption}
\usepackage{rotating}
\usepackage{tikz}
\usetikzlibrary{calc,patterns}
\usepackage{pdflscape}
\usepackage{afterpage}
\usepackage{pbox}
\usepackage{wrapfig}
\usepackage{sidecap}
\usepackage{marvosym}
\usepackage{wasysym}
\usepackage{paralist} % For inparaenum environments
\usepackage{colortbl}
\definecolor{Gray}{gray}{0.9}

\usepackage{multirow}
\usepackage{multicol}
\hypersetup{breaklinks=true}

\usepackage[]{geometry}
\usetikzlibrary{arrows,matrix,positioning}
\usepackage{soul}
\usepackage{todonotes}
\usepackage{cleveref}
\usepackage{booktabs}
\usepackage{adjustbox}
\usepackage{array}
\usepackage{xspace}

\usepackage{pifont}% http://ctan.org/pkg/pifont
\newcommand{\cmark}{\ding{51}}%
\usepackage{ifthen}

\newcommand{\trimmed}{foo}

\newcommand{\HeatmapNode}[1]{
	\ifthenelse{#1 < -200}
	{
		\renewcommand{\trimmed}{-200}
	}
	{
		\ifthenelse{#1 > 200}
		{
			\renewcommand{\trimmed}{200}
		}
		{
			\renewcommand{\trimmed}{#1}
		}
	}
	
	\ifthenelse{#1 < 0}
	{
		\pgfmathparse{round(\trimmed * -0.5)}
		\node [fill=red!\pgfmathresult] {#1};
	}
	{
		\ifthenelse{#1 = 0}
		{
			\pgfmathparse{round(\trimmed * 0.5)}
			\node [fill=blue!5] {#1\%};
		}
		{
			\pgfmathparse{round(\trimmed * 0.5)}
			\node [fill=blue!\pgfmathresult] {#1\%}; 
		}
	}
}

\begin{document}
\makeRR   % cas d'un rapport de recherche
%% \makeRT % cas d'un rapport technique.
%% a partir d'ici, chacun fait comme il le souhaite

\section{Introduction}\label{sec:intro}

%% What are privacy policies?
As of today, the main way to obtain information related to personal data collection and
processing is through \emph{privacy policies}.
Privacy policies are typically presented as textual documents
describing details such as data collection, processing, disclosure and
management.
Organizations collecting personal data (in what follows \emph{data   controllers}, or DC) commonly use privacy policies to inform individuals (in what follows \emph{data subjects}, or DS) about how personal
data is handled.
DS are often required to read these policies --- even though it rarely
occurs~\citep{jensen_privacy_2004} --- or are at least presumed to do so and to decide whether they accept the conditions.
Alternatively, giving DS the possibility of describing their own
privacy policies
%, that we denote \emph{DS privacy policies}, 
has recently gained in popularity.
This approach gives DS the time to reflect on their choices, and the possibility to consult experts and pairs.
%
%As opposed to the former approach where DS agree on DC privacy policies on-the-spot, \ie, just before starting using their services.
%
Nonetheless, privacy policies are hard to
understand~\citep{cateLimitsNoticeChoice2010}, for
DS~\citep{mcdonald_cost_2008} as for
experts~\cite{reidenberg_ambiguity_2016}.
We use \emph{DS policies} to refer to the privacy policies of
individuals, and \emph{DC policies} to denote the privacy policies of
organizations collecting personal data.

Requirements and recommendations to express privacy policies come from
different sources such as privacy regulations, authorities and
organizations.
For instance, in Europe, the General Data Protection Regulation (GDPR)~\cite{european_parliament_general_2016}
%--- the legal framework governing personal data collection and processing in Europe since May 2018 ---
requires more transparency for data processing from DC, and guidelines have been issued by the WP29\footnote{WP29 stands for Working Party 29, an European advisory board.} %, now the European Data Protection Board (EDPB).} 
to present their expectations~\cite{wp29_guidelines_2017-2}. 
These requirements are necessary for privacy policies to be compliant with the legislation.
%These requirements are all the more important that privacy policies often are the documents holding legal value when personal data processing is considered.
%
Recommendations for drafting policies have also been made by different
organizations to improve their readability.
For example, the National Telecommunications and Information Administration for mobile apps~\cite{national_telecommunications_and_information_administration_short_2013}, and the WP29 for IoT devices~\cite{wp29_opinion_2014}.
Furthermore, Data Protection Authorities (DPAs) should be able to audit data processing systems, to ensure
their compliance with the law and with the declared privacy policies.\footnote{For instance, see the decision of the ``Commission Nationale de
l'Informatique et des Libert\'es'' (CNIL), the French DPA, against
Google LLC ~\cite{cnil_cnils_nodate}.}
All these requirements and recommendations can be summarized in three
requirements:
\begin{itemize}
\item Privacy policies must be legally valid.
\item Privacy policies must be understandable by all parties.
% (including lay-users).
\item Privacy policies must be
  %effectively
  enforceable and auditable in data processing systems.
  % through auditable mechanisms.
\end{itemize}

%%% note(raul): this paragraph seems to be more confusing than clarifying
%
% Privacy regulations impose the requirements on the content that
% privacy policies must include, and how data must be
% processed.\raul{Removed for the deliverable but for the SoK we
% should add something.}
%
% Organisations and authorities provide guidelines for DC to make
% privacy policies accessible to lay-users.
%
% Data processing by DC must match the requirements of the privacy
% policies.
%
% In other words, DC should enforce the privacy policies.

%%% Requirements are met by different types of privacy policies separately
Existing methods to express privacy policies address some of these
requirements, but not all of them.
Different methods have arisen from different needs, and they target
different audiences --- from expert to lay-users.
For instance, legal privacy policies are often written as long and complex documents which are necessary in court, but that are not easy to understand for lay-users.
% --- on the rare occasions they attempt to them \citep{jensen_privacy_2004}.
%
As an attempt to simplify these legal documents, organizations work on
summarized versions of privacy policies or use visual aids to help users understand the risks of having their data collected.
%
% Another way forward in this direction is the use of techniques for
% extracting relevant information from existing privacy policies through
% Natural Language Processing (NLP) \citep{wilson_demystifying_2016}.
% %
% However, it is not powerful enough to be applicable to existing
% privacy policies yet.
%
Ensuring that data is processed according to the requirements in
privacy policies is not an easy task either.
Some work --- coming mostly from academia --- proposes an alternative
format for privacy policies that can be read by computers.
These solutions aim at bridging the gap between the textual legal
requirements and their enforcement in the underlying system.
Furthermore, some of these proposals are equipped with auditing tools
which facilitate, for DPAs or DS, verifying that no violations of a
privacy policy have occurred.
Unfortunately, these solutions are not widely used.

In this work, we analyze the state-of-the-art methods on expressing
privacy policies.
We provide a comprehensive picture of existing proposals in order to
identify gaps and challenges.
In doing so, we have studied the three main ways to
express privacy policies: natural language, graphical and
machine-readable; which we call the \emph{facets} of privacy
policies.
Each of these facets have mostly arisen from different
communities.
Natural language comes from law experts, graphical from organizations and privacy advocates, and machine-readable from academics.
Consequently, the content of this paper contextualizes knowledge often
studied within different communities that have been working on similar
issues, but with different objectives.
Unsurprisingly, each facet mainly provides benefits to the
specific community it was defined in.
Therefore, we take the insights of our study to explore how the different facets of privacy policies can complement each other.
This synergy includes the benefits of each facet and minimize their limitations.
We discuss guidelines and challenges of combining all facets in a
single policy, which we denoted as \emph{multi-faceted privacy
  policies}, and study recent efforts in this direction.
% 
%Hence, we propose a new type of privacy policies combining aspects from all facets, which we denote \emph{multi-faceted privacy policies}.
%

More concretely, in bringing the above ideas to the forefront, our
contributions are:
%
% In bringing these challenges~\victor{Which challenges?} to the forefront, our contributions are:
\begin{enumerate}
	\item A categorization of existing work in each facet according to a privacy taxonomy (introduced in Section~\ref{sec:methodology}), and the specific features of each facet.
\item An in-depth study of the existing facets of privacy policies:
  \begin{inparaenum}[i)]
  \item privacy policies expressed in natural language (Section~\ref{sec:natural_language}),
  \item \gpp (Section~\ref{sec:graphical}), and 
  \item \mrpp (Section~\ref{sec:machines}).
  \end{inparaenum}
  For each facet, we provide an overview of:
  % \begin{inparaenum}[i)]
  % \item
  its content;
  % \item
  the available tools;
  % \item
  its benefits; and
  % \item
  its limitations.
  % \end{inparaenum}
\item Insights from the study on future research on designing privacy policies (Section~\ref{sec:perspectives}).
\end{enumerate}
%Section~\ref{sec:methodology} describes the  systematization methodology.
%In sections \ref{sec:natural_language}-\ref{sec:machines} we study the different facets.
%We present the insights of our study in Section~\ref{sec:perspectives}.
Section~\ref{sec:conclusion} discusses related work and concludes the paper.

\section{Systematization Methodology}
\label{sec:methodology}

For each facet, we study the content of the privacy policies, the
available tools, benefits and limitations.
The literature of privacy policies across the different facets is vast.
An exhaustive analysis of all works is unfeasible and undesirable.
Instead, we focus on highlighting representative work; hence we do not
provide an all-encompassing reference to every related work.

We categorize the content in each facet according to a taxonomy of
privacy policies.
Several taxonomies have been proposed to categorize the content of
privacy policies~\cite{solove_taxonomy_2005, wilson_creation_2016,
	janczewski_assessing_2018}.
We use a slight variation of~\cite{wilson_creation_2016}.
This taxonomy is appropriate for our purposes for two main reasons:
\begin{inparaenum}[1)]  
	\item it was devised for a wide range of privacy policies and therefore reflects their content across facets; and
	%
	% \item it was devised according to existing \nlpp and therefore reflects their content; and
	\item it encompasses most requirements of the current legislations and guidelines, such as the GDPR, the Fair Information Practice Principles
	(FIPPs)~\cite{federal_trade_commission_fair_2000} in some
	cases, and the California Consumer Privacy Act of 2018 (CCPA)~\cite{stateofcaliforniaAssemblyBillNo2018}.
%	~\footnote{We discuss below the relevance of legislations with respect to the content of policies.}
\end{inparaenum}
%
%This taxonomy was also used in Polisis \cite{harkous_polisis_2018} as discussed in Section~\ref{subsec:analysis}.

\subsubsection*{Taxonomy}

Wilson \etal~\cite{wilson_creation_2016} proposed a taxonomy for
privacy policies composed of the following items:\footnote{We denote
	\textit{item} a piece of information provided in a privacy policy.}
\begin{inparadesc}
	\item ``\textit{First Party collection}: How and why a service provider collects user information'',
	\item ``\textit{Third Party collection}: How user information may be shared with or collected by third parties'',
	\item ``\textit{Access, Edit, Delete}: If and how users may access, edit, or delete their information'',
	\item ``\textit{Data Retention}: How long user information is stored'',
	\item ``\textit{Data Security}: How user information is protected'',
	\item ``\textit{Specific Audiences}: Practices that pertain only to a specific group of users (e.g., children, Europeans, or California residents)'',
	\item ``\textit{Do-Not-Track}: If and how Do Not Track signals for online tracking and advertising are honored'',
	\item ``\textit{Policy Change}: If and how users will be informed about changes to the privacy policy'',
	\item ``\textit{Other}: Additional sub-labels for introductory or general text, contact information, and practices not covered by the other categories'', and
	\item ``\textit{Choice Control}: Choices and control options available to users''.
\end{inparadesc}

Our variation of Wilson \etal's taxonomy accommodates it to the purposes of our study, but does not change the content.
Concretely:
%, the differences are:
\begin{inparaenum}[i)]
	\item we use \textit{DS rights} to denote \textit{Access, Edit,
		Delete} and \textit{Choice Control} as they relate to DS rights in
	the sense of the GDPR (see Chapter III of the GDPR);
	\item subsume \textit{Specific Audiences} and \textit{Do-Not-Track} under
	\textit{Other} as they are occasional items; and
	\item we observe that a legal requirement is missing in the taxonomy,
	even though it is often found in \nlpp: the \textit{legal basis} of
	processing, we will therefore add it to our taxonomy.
\end{inparaenum}
These differences better accommodate recent regulations (for DS rights and Legal basis) and practices (for Other).
%The differences are motivated by concern of a mapping with recent regulations (for DS rights and Legal basis) and practices (for Other).
%
Table \ref{tab:content} summarizes the taxonomy.
\Cref{sec:nl_content} provides a detailed explanation of each taxonomy item together with illustrative examples.

\begin{table}[!ht]
  \centering
  \scalebox{.8}{
	\begin{tabular}{p{2.3cm}p{7cm}ccccc}
		% \hline
		\multicolumn{1}{c}{Taxonomy item} & \multicolumn{1}{c}{Description} & \multicolumn{1}{c}{GDPR} & \multicolumn{1}{c}{FIPPs} & \multicolumn{1}{c}{CCPA} & \multicolumn{1}{c}{HIPAA$_a$} & \multicolumn{1}{c}{COPPA$_b$} \\ %\hline
		\midrule
		\makecell[cl]{First Party \\ collection} & Type of data collected, purpose and collection mode. & \cmplt & \cmplt & \prtl & \cmplt & \cmplt \\ %\hline
		
		\rowcolor{Gray}
		\makecell[cl]{Third Party \\ collection} & Type of data collected, purpose and collection mode for third parties. & \cmplt & \cmplt & \prtl & \prtl & \cmplt \\ %\hline
		
		Legal basis & Ground on which is determined the lawfulness of processing. & \cmplt & \prtl & \Circle & \Circle & \Circle \\ %\hline
		
		\rowcolor{Gray}
		DS rights & Rights of the DS, \eg, right to access, to rectify, to port or erasure. & \cmplt & \Circle & \prtl & \cmplt & \cmplt \\ %\hline
		
		Data Retention & Duration of data storage & \cmplt & \Circle & \Circle & \Circle & \prtl \\ %\hline
		
		\rowcolor{Gray}
		Data Security & Modalities of protection of data, \eg, encrypted communication and storage. & \prtl & \cmplt & \Circle & \cmplt & \prtl \\ %\hline
		
		Policy Change & Modalities of notification for policy changes. & \prtl & \Circle & \Circle & \Circle & \Circle \\ %\hline
		
		\rowcolor{Gray}
		Other & Other items such as identity of DC, information related to Do-Not-Track, to children \ldots & \cmplt / \prtl & \cmplt / \prtl & \cmplt / \Circle & \prtl & \cmplt \\ %\hline
		\bottomrule
	\end{tabular}}
	\caption{Summary of our taxonomy with the legal requirements of items. We use \cmplt\ to denote \textit{Required explicitly}; \prtl\ to denote \textit{Addressed but not required}; and \Circle\ to denote \textit{Absent}. The subscript $_a$ means that HIPAA only considers health data. The subscript $_b$ means that COPPA only considers personal information from children, and notice must be addressed to parents.
		\vspace*{-.75cm}}
	\label{tab:content}
\end{table}

\section{Natural language privacy policies}
\label{sec:natural_language}
%Privacy policies are typically expressed in natural language: it has been the dominant format of privacy policies since their widespread adoption~\footnote{This widespread adoption can be traced to 1981, when the \textit{Convention for the protection of individuals with regard to automatic processing of personal data} --- a European treaty recommending the use of privacy policies --- was adopted.}, and we will therefore denote them \textit{\nlpp}. 

Most legislations require notices expressed in natural language to inform DS about the collection and processing of their personal data:
%~\footnote{For example, the GDPR mentions a list of information to provide where personal data is collected in Art. 13 and 14.} 
the use of natural language is necessary to ensure that the policy has legal value (\eg,~\cite[Art. 13 \& 14]{european_parliament_general_2016}).
The ways these documents can be authored --- \ie, drafted automatically or written manually --- and the manners to assist their authoring can vary greatly. 
%There are many ways to express privacy policies in natural language. 
We present the content expressed by \nlpp~in Section \ref{sec:nl_content}, the tools used to assist their authoring and to analyze existing \nlpp~in Section \ref{sec:nl_tools}, the benefits in Section \ref{sec:nl_benef}, and the limitations in Section \ref{sec:nl_limit}.

\subsection{Content}
\label{sec:nl_content}
%\paragraph{Natural language can express anything}
\Nlpp~are familiar to the public as they have been adopted by a large range of online services such as social networks, file hosting services, mobile applications, \etc
Because these privacy policies are expressed in natural language, they are not restricted in terms of content; they cover all the items of our taxonomy (see~\Cref{sec:methodology}). 
%
% Several taxonomies have been proposed to categorized the content of privacy policies~\cite{solove_taxonomy_2005, wilson_creation_2016, janczewski_assessing_2018}.
% %
% In this paper, we use a small variation of~\cite{wilson_creation_2016}.
%
% We will use this taxonomy for the rest of the document.
%
In the following we examine the taxonomy items in detail, and discuss what
legal requirements appear explicitly in the GDPR, the FIPPs, or in the CCPA.

We focus on the requirements of the GDPR, the FIPPs, and the CCPA as they are the three main texts (legislations or guidelines) that determine the content required when informing DS of data collection and processing.
%~\footnote{See Appendix~\ref{app:content_req} for the details.}
%
The GDPR is the text regulating personal data collection and processing in the EU, and many countries consider it since it has an extraterritorial scope.
The FIPPs are guidelines designed by the United States Federal Trade Commission's (FTC).
They represent common principles regarding fair information practices.
However, they have been criticised by scholars~\cite{cateFailureFairInformation}, and have not been updated in decades.
The CCPA is more recent (2018), it is a privacy enhancing and consumer protection bill for Californians.
It has been deemed ``the strongest privacy controls of any state in the U.S''~\cite{CaliforniaNewPrivacy}.
%It has been referred to ``Almost GDPR in the US''~\cite{CaliforniaNewPrivacy}.
%
We also consider other widely-known regulations, such as health data for HIPAA~\cite[Notice and Other Individual Rights]{united_states_congress_health_1996}, or children for COPPA~\cite[§ 312.4]{federal_trade_commission_childrens_2013}).
%, but they have a restricted scope and cannot compare to the broad requirements of the three main other texts.~\footnote{However, their requirements can be combined, \eg, data collection and processing for a Californian child falls under both CCPA and COPPA.}
%
%In the following we examine the taxonomy items in detail.
%, illustrated with examples from existing \nlpp~such as Facebook~\cite{facebook_facebooks_nodate}, Twitter~\cite{twitter_twitter_nodate}, Dropbox~\cite{dropbox_dropbox_nodate}, Netflix~\cite{netflix_netflix_nodate}, and Google~\cite{google_privacy_nodate}.

%\vspace*{-.5cm}

\paragraph*{First Party Collection}
The most common item in \nlpp is the first party collection, which describes \textit{what} data is collected, \textit{why} it is collected, and sometimes \textit{how}. 
The type of data ranges from generic to more precise assertions, \eg, respectively \textit{we collect your data} and \textit{your email address is collected}.
%\paragraph{Mundane data}
%Common types of data collected can be the name of the DS, an email address, geolocation, messages, \etc; or the social graph, more specifically to social networks. 
%As an example, Facebook collects ``Networks and connections. [...] information about the people, Pages, accounts, hashtags and groups you are connected to and how you interact with them across our  Products[...]. We also collect contact information [...] (such as an address book [...].''
%\paragraph{Cookies}
Cookies~\cite{degelingWeValueYour2019}
%~\footnote{Small pieces of data sent from a website and stored on the DS's computer by the DS's web browser~\cite{degelingWeValueYour2019}.} 
often have a distinct treatment, most likely because they are often collected by websites: it is common to find a dedicated paragraph for their management.
% in a natural language privacy policy. 
%\paragraph{Geolocation}
Location information is often treated in a separate section as well because it can be collected from different sources --- mobile applications, web browsers --- or inferred from metadata --- such as IP addresses. 
%For instance, Twitter's privacy policy states: ``Location Information: We require information about your signup and current location, which we get from signals such as your IP address or device settings, to securely and reliably set up and maintain your account and to provide our services to you.''
%\paragraph{Purposes}
The purpose of processing often goes hand in hand with the type of data collected.
%\paragraph{Advertisement}
%It is possible to find among the purposes, \textit{marketing} and \textit{advertising}, which are prevalent in \nlpp.
%\paragraph{Analytics}
%Analytics is often mentioned as a purpose to improve the functioning of services, to provide a better overview of what is actively used or not in a service, or to automatically retrieve malfunctions.
%\paragraph{Security}
%Data can also be collected for security reasons: to remove illegal or harmful content, or to prevent payment fraud.
%\paragraph{Research}
%Certain services collect data for research, and this broad purpose can be exempt of some constraints for the definition of a more concrete research purpose~\cite[Recital 159]{european_parliament_general_2016}.
%~\footnote{See Recital 159 of the GDPR.}
%\paragraph{Core functioning}
%DC can conduct data collection to operate a service: Facebook for instance mentions ``Provide, personalize and improve our Products'' as a purpose of processing, and this is a reason often put forward for data processing.
%\paragraph{Collection mode}
It is also possible to find the collection mode in some \nlpp: whether the data is collected automatically, by manual input of DS, or by any other means.

Informing about the type of data, the purpose of processing, the recipients and the means of collection is required by the GDPR and the FIPPs.
The CCPA gives the right to request first party collection, but does not require it.
HIPAA requires to inform of ``the ways in which the covered entity may use and disclose protected health information''.
COPPA requires to inform of ``what use, if any, the operator will make of the personal information collected''.

\paragraph*{Third Party Collection}
Third Party Collection is a common item in \nlpp, and it is therefore usual to find the third-parties to whom data will be transferred: they can be advertisers, or other business partners. 
The notion of \textit{sharing} can also refer to other DS and subsidiary companies. 
It is usually composed of the same content as First Party Collection.
%, \textit{i.e.} type of data and purpose. 
%For instance, Dropbox declares in its privacy policy: ``Dropbox uses certain trusted third parties (such as providers of customer support and IT services) to help us provide, improve, protect and promote our Services. 
%These third parties will only access your information to perform tasks on our behalf in compliance with this Privacy Policy, and we'll remain responsible for their handling of your information per our instructions. 
%For a list of trusted third parties that we use to process your personal information, please see our FAQ.'' 

Informing about third party collection is required by both the GDPR, the FIPPs, and COPPA.
%~\footnote{\textit{Recipient} encompasses first and third party collection in the GDPR .}
The CCPA gives the right to request categories of third parties, but does not require it except if data is sold to those third parties.
HIPPA considers it implicitly (see~\cite[Who is Covered by the Privacy Rule]{united_states_congress_health_1996}).
%~\footnote{``The Privacy Rule covers a health care provider whether it electronically transmits these transactions directly or uses a billing service or other third party to do so on its behalf''~\cite[Who is Covered by the Privacy Rule]{united_states_congress_health_1996}.} 

\paragraph*{Legal basis}
Legal basis (or legal ground) is regularly found as a complement of the purpose of processing.~\footnote{Art. 13(1)(c) of the GDPR requires "The purposes of the processing for which the personal data are intended \textbf{as well as} the legal basis for the processing." (Highlights from authors)}
%\paragraph{Consent}
A common legal basis for processing is consent, which consists, for DC, in retrieving an authorization from DS to legally collect their data. Consent has to be informed and specific under the GDPR~\cite[Recital 32]{european_parliament_general_2016},
%~\footnote{"Freely given, specific, informed and unambiguous indication of the data subject's agreement" in Recital 32.} 
and it still is often used as a legal basis. DC might consider the reading of their \nlpp~as a proper consent, without questioning the conditions to obtain consent~\cite{forbrukerradet_deceived_2018}.
%\paragraph{The rest}
Other legal basis can be found in \nlpp, such as the necessity for the performance of a contract, compliance with legal obligations, protection of DS's vital interests or public interest, and the legitimate interests of a DC. These legal basis are listed in the GDPR~\cite[Art. 6]{european_parliament_general_2016},
%~\footnote{See Art. 6.} 
and major stakeholders generally consider cumulatively either all of them, 
% --- such as Facebook which combines all possible legal bases ---
  or a large subset.
%   --- \eg, Netflix's policy considers all of them except public interest.

Informing about the legal basis is required by the GDPR, and not explicitly by the FIPPs which requires informing ``whether the provision of the requested data is voluntary or required, and the consequences of a refusal to provide the requested information''.
Legal basis information is not required by CCPA, HIPPA, and COPPA.

\paragraph*{DS Rights}
%\paragraph{Opt-out}
DS can exercise rights regarding their data, and \nlpp~now often mention the rights to access, rectify, port and erase data, likely due to the influence of the GDPR.
%~\footnote{These rights are explicitly mentioned in the GDPR.} 
%As an example, Google's privacy policies mentions: ``You can export a copy of your information or delete it from your Google Account at any time''. 
DS rights can be seen more restrictively as possibilities to opt-in or opt-out.
%~\footnote{Note that opt-out is now illegal in Europe.} 
Thus \nlpp present how to subscribe or unsubscribe to specific services.

Informing about DS rights is required by the GDPR, HIPAA and COPPA, but not by the FIPPs.
In CCPA, only the right to opt-out is explicit (other rights are ensured but Californians do not have to be explicitly informed of them in a privacy policy).
% or in a privacy notice).

\paragraph*{Data Retention}
\Nlpp~often describe the period during which personal data will be stored. It can be a fixed value --- \eg, \textit{30 days after data collection}, or variable --- \eg \textit{as long as your account is active}. It often comes with the type of data, the purpose, and the legal basis of processing.

Informing about the retention time is required by the GDPR.
COPPA addresses it but do not make it mandatory to inform about it.
FIPPs, CCPA and HIPPA do not require it.

\paragraph*{Data Security}
DC regularly explain in their policies how data is stored, if its communication is secured or its storage encrypted. 
%As an example, Netflix's privacy policy claims: ``Security: We use reasonable administrative, logical, physical and managerial measures to safeguard your personal information against loss, theft and unauthorized access, use and modification. 
%These measures are designed to provide a level of security appropriate to the risks of processing your personal information.''

Informing about the security of data is required by the FIPPs, but not by the GDPR although it mentions that ``Personal data should be processed in a
manner that ensures appropriate security and confidentiality of the personal data''~\cite[Recital 39]{european_parliament_general_2016}.
%~\footnote{See Recital 39 of the GDPR.}
Similarly, COPPA states that ``The operator must establish and maintain reasonable procedures to protect the confidentiality, security, and integrity of personal information collected from children.''
HIPPA requires to inform of the ``entity's duties to protect privacy''.
CCPA does not require to inform of this item.

\paragraph*{Policy Changes}
%\paragraph{Information of changes}
The modalities of notification in the case of a change in the privacy policy can also be observed.
Notification is usually by email or within the service's interface, in some cases notifications are sent by regular mail or by phone.
This item is not required by the GDPR, the FIPPs, CCPA, HIPPA, or COPPA.

\paragraph*{Other}
%\paragraph{Contacting the DC}
We subsume the identity and contact of DC, requirements towards specific audiences such as children, and Do-Not-Track under this item.
The DC usually provides its identity, as well as its contact details if the DS has to lodge a complaint.
%\paragraph{Children}
In many legislations~\cite[Recitals 38 \& 58]{european_parliament_general_2016}~\cite{federal_trade_commission_childrens_2013}
%,~\footnote{In particular Recitals 38 and 58 of the GDPR, and the Children's Online Privacy Protection Act (COPPA).} 
children have specific considerations. 
As a result, many \nlpp include a dedicated section, even if it is only to mention that personal data of children under thirteen is not collected without parental consent.

The identity of DC is required by the GDPR, the FIPPs, and HIPAA, but not Do-Not-Track.
As for requirements for children, they are specifically addressed by these legislations.
CCPA requires more specifically to ``Make available to consumers two or more designated methods for submitting requests for information'', and does not address data collected from children.
COPPA requires to inform of the identity and contact of DC, and targets data collection related to children.

\subsection{Tools}
\label{sec:nl_tools}
Several solutions exist to assist in the authoring of \nlpp, offering different levels of automation. 
%ranging from the least to the most automated ones. 
We denote these tools \textit{authoring tools}. We distinguish \textit{templates}, \textit{generators}, and \textit{retrievers}. 
%While natural language does not limit the expressiveness of privacy policies, 
Authoring tools are often tailored to websites and mobile app owners, and are constrained in terms of content.
%
%In addition to the authoring tools presented above, 
Furthermore, there exist \textit{analysis tools}: software able to parse and analyze \nlpp, in order to produce a machine-readable or a graphical
version of a policy.

\paragraph*{Templates and generators} Tools such as Docracy~\cite{docracy_open_2012}, Termsfeed~\cite{termsfeed_generic_nodate},
SEQ Legal~\cite{taylor_privacy_2012}, and 3DCart~\cite{3dcart_create_nodate}
provide a \emph{fill-in-the-gap} form, where the author writes
appropriate terms in the fields.
We refer to them as \emph{templates}.
These tools guide the authoring process, but do not provide additional functional such as validation of the input data.
%The redundancy is not taken into account, and no verification can be made regarding the validity of the terms written. 
It is not possible, for instance, to check if the email of a service owner is valid. 
\emph{Generators} comprise more complex tools where authors input data only once regarding relevant fields of a privacy policy, and the tool automatically generate the privacy policy.
%
% More complex tools differ from templates by inputting information in a form using software components.
% We denote them \emph{generators}.
%They usually do not allow the possibility to describe the type of data collected, nor its use: their expressive power often cannot provide all the items required for legal compliance --- at least in EU. Their level of automation is low, and the result is easily prone to errors. 
%\paragraph*{Generators}
%The author inputs information only once. 
Most generators do not allow incorrect data: email addresses without \verb+@+ are highlighted, and the author cannot go further in the process and generate the policy. 
Generators also give the option of expressing the same policy according to different legal contexts.
%Legislation varies between countries. 
%A generator can 
%overcome this issue, and 
%propose text corresponding to the applicable legal framework. 
For instance, \url{privacypolicies.com} offers clauses specific to the GDPR or COPPA. %for an additional cost, if required. 
A set of what we denote \textit{light} generators have a restricted set of parameters, while another composed of more \textit{detailed} generators offer a choice between a policy tailored to websites or mobile applications, with a more exhaustive list of items~\cite{kolter_generating_2009}.
Privacy Policy
Generator~\cite{privacy_policy_generator_privacy_nodate}, Privacy
Policy Online~\cite{delpercio_privacy_nodate}, or
GetTerms~\cite{getterms_getterms.io_nodate} are examples of the
former set, and
\url{PrivacyPolicies.com}~\cite{privacypolicies.com_privacy_nodate}
and
\url{FreePrivacyPolicy.com}~\cite{freeprivacypolicies.com_free_nodate}
of the latter.

\paragraph*{Retrievers}
Retrievers, such as those offered by Miao \cite{miao_privacyinformer_2014}, Apolinarski \etal \cite{apolinarski_automating_2015} and Yu \etal \cite{yu_autoppg_2015}, automatically extract relevant information from code of mobile application using static code analysis or user behavior analysis.
For instance, in \cite{apolinarski_automating_2015} the authors analyze sharing behavior when using online collaboration tools.
These prototypes work on Android applications, in which personal data management is structured around the concept of permissions~\cite{google_android_nodate}. 
These permissions define the type of data accessible by an application, such as contacts, content of text messages, or Wi-Fi management. 
A \textit{retriever} analyses those permissions, and interprets them according to well-defined rules to author a natural language privacy policy. 
In that case, an author does not necessarily have to input any information in addition to the code: the \textit{retriever} can parse the name of the DC, the permission requested by a service or the third-party libraries, and can convert this information into natural language. 
However, \textit{retrievers} are often tailored to a specific solution --- mobile applications in most cases --- and could thus be difficult to implement in other ecosystems. 
Furthermore, they cannot automatically retrieve certain information, such as the purpose of collection or the retention time. 
%\textit{Retrievers} reach the highest level of automation among authoring tools. 

\paragraph*{Analysis tools}
\label{subsec:analysis}
Analysis tools have been developed for over a decade.
They focused on using Natural Language Processing or Information Extraction~\cite{costante_machine_2012} to parse \nlpp.
An early work has been conducted by Brodie~\etal~\cite{brodie_empirical_2006}.
The Usable Privacy Project lead by Sadeh~\cite{sadeh_usable_2013} further investigated the automated classification of privacy policies.
Within this project, Ammar~\etal~\cite{ammar_automatic_2012} conducted a pilot study for automatic text categorization.
The Usable Privacy Project also developed a website privacy policy corpus~\cite{wilson_creation_2016}, which will later be notably used by Polisis~\cite{harkous_polisis_2018}.
Zimmeck~\etal~\cite{zimmeck_privee_2014} devised a hybrid solution combining machine learning classifiers (association rules) with crowdsourcing.
Analysis tools achieve an accuracy averaging around 80\%, which has not significantly improved since the first attempts.
\\

The content produced by most authoring tools does not have legal value. 
Those tools do not provide legal advice, but rather general guidelines for policy authoring. 
These guidelines may be sufficient, but their legal validity is not guaranteed and should be verified by a lawyer. 
As an example, Iubenda advertises for its ``Attorney-level compliance''~\cite{iubenda_features_nodate}, but advocates for a professional legal consultancy
%: ``Nothing can substitute a professional legal consultancy in the drafting of your privacy policy''
~\cite{iubenda_terms_nodate}, and do not guarantee conformity with the law, which they claim ``only a lawyer can do''. In other words: DC are responsible for the compliance with the law. 
They have to ensure that their privacy policies address all legal requirements, and to enforce the claims made in their policies.

\subsection{Benefits}
\label{sec:nl_benef}
%\paragraph{What can they guarantee?}
%The main benefit of \nlpp~is their legal value. 
%; but also compliance towards processing --- DC should act according to their privacy policies, this second aspect of compliance is out of the scope of this paper. These texts should \textit{theoretically} guarantee legal compliance, but rarely do in practice. 

\paragraph*{Legal value}
\label{par:legal_value}
\Nlpp are the only type of privacy policies with legal value as it is
the standard format for legal texts.~\footnote{It does not however mean than other facets do not have legal value, rather that \nlpp are mandatory.}
Most legislations require DC to provide a lawful statement detailing the processing of personal data, and \nlpp~often aim to fulfill this obligation.
%~\footnote{For instance, DC, in order to be compliant with the GDPR, must inform of the following: their identity and contact, the type of data collected, its purpose and its legal basis for processing, the recipient of data, the third-parties involved, the retention time, and the rights of the DS.}
%
Lawyers rely on natural language to evaluate whether privacy policies
are correctly drafted: these policies contain all details to
determine whether there has been a violation.
Also, lawyers use these policies to check compliance with data protection regulations.
%, such as the GDPR.
%
However, a document holding legal value is not necessarily compliant with the law.
For instance, lawyers or DPAs may check that all items required by the legislation are provided to DS, and auditors can check that data processing is performed according to the policy.
Legal compliance is twofold: with respect to the information
requirements, and with respect to the actual processing.

\subsection{Limitations}
\label{sec:nl_limit}
\paragraph*{Ambiguity}
\Nlpp can be
ambiguous~\cite{reidenberg_automated_2016}, as they may be interpreted
in different ways. Reidenberg
\etal~\cite{reidenberg_automated_2016,reidenberg_disagreeable_2015}
presented privacy policies to privacy experts, law and policy
researchers, who were ultimately unable to agree on some aspects of the
policies.  They proposed a crowd-sourcing annotation to tackle this
issue, but admit that it would only provide a partial solution.
This ambiguity is mainly due to the fact that a statement in natural
language can be interpreted in different ways.
Ambiguity has a direct impact on the understanding, the enforcement,
and the auditability of privacy policies.
Ambiguity is also an explanation of the inaccuracy of analysis tools described in Section~\ref{subsec:analysis}.

\paragraph*{Understanding} McDonald and Cranor~\cite{mcdonald_cost_2008} showed that it would take 200 hours a year for an average US citizen to read all the \nlpp~of the online services she used. This is clearly impractical, and thinking that DS read privacy policies before using a service is a \textit{fictio juris}. All the more, nowadays, it seems highly inconvenient to spend a significant amount of time before using an online service.

\paragraph*{Enforcement \& auditability}
Because they are currently ambiguous, \nlpp~are difficult to enforce:
natural language lacks precise semantics, making it difficult to
decide how data must be processed by the underlying system.
Likewise, \nlpp can also be hurdles to auditing: it can be difficult for an
independent authority to compare stated and existing processing.

%These two limitations --- ambiguity and unreadability --- make \nlpp~hard to rely on, but are not intrinsic. These limitations should rather be seen as pitfalls, which are unfortunately also part of the current authoring tools: these tools mostly reproduce the mistakes of existing privacy policies and are unable to overcome these issues. But authoring tools could produce unambiguous and concise privacy policies if they consider only a restricted vocabulary or set of clauses (as the Design your Privacy project proposes~\footnote{\url{https://www.privacytech.fr/design-your-privacy/}[fr]}). 

%\noindent
%\textbf{Summary.~}

\subsection*{Summary}
\Nlpp~are the most used medium to express privacy policies, and the tools used to assist their production can be categorized into \textit{templates}, \textit{generators}, and \textit{retrievers}. 
These tools are often tailored to specific solutions, such as website or mobile applications, therefore restricting their scope. Analysis tools are not accurate enough to be trusted blindly. \Nlpp~are necessary for legal compliance, but suffer in practice from ambiguity and understandability.

\section{Graphical privacy policies}
\label{sec:graphical}

In the previous section, we analyzed \nlpp.
They are necessary for legal compliance, but they can mislead DS when they attempt to read them, as they are often difficult to understand.
%Although they are the most common way to express privacy policies, they are not the only one.
As a consequence, other representations focused on DS understanding have been devised: privacy policies can also be expressed with graphical (in the broad sense) representations, that we denote \textit{\gpp}.
\Gpp~cover icons sets and standardized notices as well as solutions providing additional information, such as warnings or judgments, sometimes combined with simple text~\cite{rossiWhenDesignMet2019}.
Note that \gpp are rarely specifications, rather additional representations of \nlpp.
For instance, Android Permissions~\cite{google_android_nodate} represent privacy policies combining icons and text according to XML specifications. 
\Gpp~often come from privacy advocates, but this is not only the case, notably since the WP29 explicitly mentioned icons as appropriate to convey privacy notices in their guidelines for transparency~\cite{wp29_guidelines_2017-2}. 
Privacy notices are means to inform DS, and are often understood as what we denote \gpp.
Therefore, we include them in our study. 
% We review in this section these means to express privacy policies graphically.
% In particular, 
We categorize each work based on:
\begin{inparaenum}[i)]
	\item the elements in the taxonomy presented in
	Section~\ref{sec:natural_language} that it captures;
	\item its features, whether it is made of icons, complementary text, or variants of those; and
	\item the intended audience of the language, \eg, DS or DC.
\end{inparaenum}
Table~\ref{tab:framework_graphical} in Section~\ref{subsec:gpp-sum}
summarizes our study.

\subsection{Content}
\label{sec:graph_content}
% As discussed in Section \ref{sec:natural_language}, \nlpp~present specific items, notably due to the influence of legal requirements.
Based on their content, \gpp can be divided into three main types:
\emph{icons}, \emph{standardized notices}, and \emph{rating solutions}.
This distinction emerged from an empirical analysis of the solutions studied, combined with impactful insights such as~\cite{cranor_necessary_2012}.
\Gpp based on icons intend to express the content of privacy policies, for DC and DS policies.
Some of these icons try to cover all the items of the taxonomy introduced in Section~\ref{sec:natural_language}.
Other \gpp aim to express the same content as \nlpp, but in a standardized and often comparable manner.
%
%Most of the solutions belonging in this category pertain to P3P.~\footnote{P3P is an obsolete set of specifications for websites to declare their \DCPs. See Section~\ref{sec:machines}.}
%
Some \gpp provide rating information concerning certain aspects of privacy policies such as the transparency level or potential risks.
These solutions were not devised to meet the same requirements as \nlpp, the content of these \gpp in that respect is often restricted.
%
%However, they highlight important information for \dss.
%
In the following, we describe the content of \gpp according to the
elements of the taxonomy and their
type.
% (icons, standardized, or rating).
%and whether they are intended to be used by DS or DC.

\subsubsection*{Sets of icons}
\label{sec:graph-icons}
The content of \gpp~reviewed in this section lies in their icons, and sometimes in the simple explanations that comes with them.
As an example, the symbol @ can represent collection of an email address, and a stylized calendar \includegraphics[height=\fontcharht\font`\B]{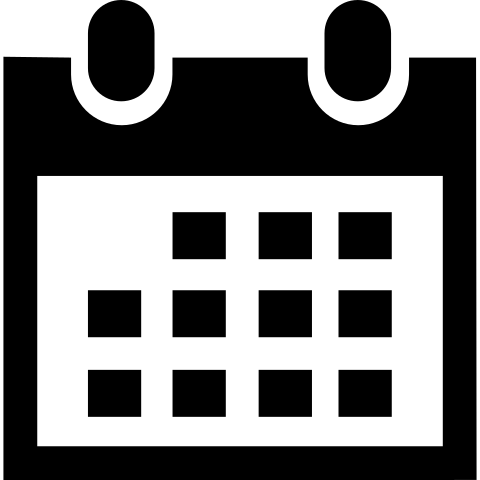} can represent retention time. 
But certain items are harder to express graphically. 
For instance, describing the legal basis of processing with the help of icons can easily be mistaken, and can mislead the intended audience instead of simplifying the understanding. 
%In practice, \gpp~have a restricted set of icons, and express specific items (seen in Section \ref{sec:nl_content}).
%Sets of icons initially emerged from academia, but they were quickly adopted by privacy advocates. 
%A notable solution comes from the business sector.

Usually, sets of icons (see~\cite{camenisch_privacy_2011, egelman_is_2015}) do not sufficiently express the items presented in the taxonomy.
Instead, they express specific items.
A solution such as Privicons~\cite{konig_privacy_2012} focuses on informing mail correspondents of how the data should be handled instead.
% (see Figure \ref{fig:privicons}).
Other solutions such as~\cite{rundle_international_2006} include icons for selling, and second-use of data, but the type of data cannot be specified.
%; or Egelman~\etal~\cite{egelman_is_2015} focuses only on the type of data collected --- voice, gesture, image --- and its purposes --- detection of gender, emotion, language. 
Recently, Rossi and Palmirani~\cite{rossiDaPISOntologyBasedData2019} proposed a Data Protection icon set, named DaPIS.
The interesting features of their approach is that they based the icon set on an ontology named PrOnto, and they tested their set in order to refine it.
Moreover, it stands out by emphasizing items recently introduced in the legislations, such as DS rights or legal bases for processing.
% (see Figure~\ref{fig:dapis}).
However, it does not consider the type of data.

\begin{figure}[!ht]
	\centering
	\begin{subfigure}[h]{0.2\textwidth}
		\centering
		\includegraphics[scale=.1]{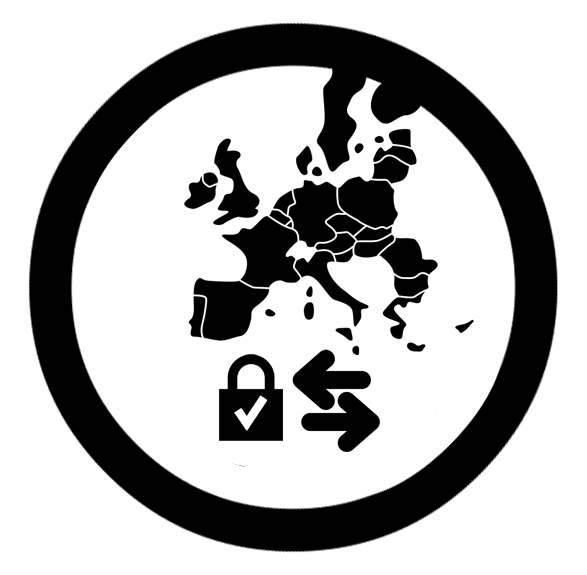}
		\caption{\centering \footnotesize UE transfer adequacy}
	\end{subfigure}
	~
	\begin{subfigure}[h]{0.2\textwidth}
		\centering
		\includegraphics[scale=.1]{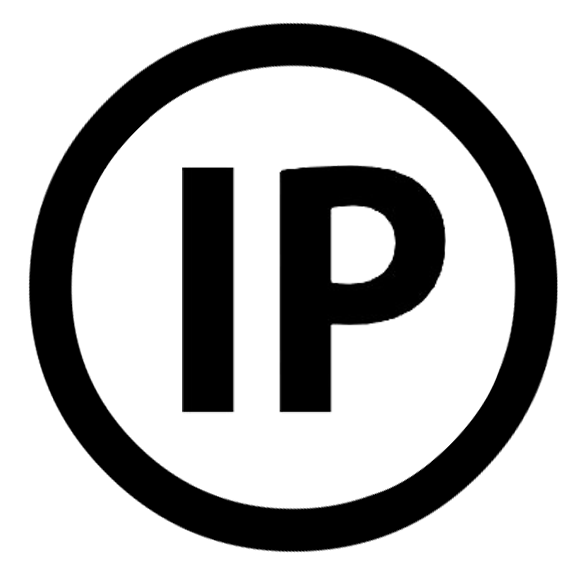}
		\caption{\centering \footnotesize Connection data}
	\end{subfigure}
	~
	\begin{subfigure}[h]{0.2\textwidth}
		\centering
		\includegraphics[scale=.1]{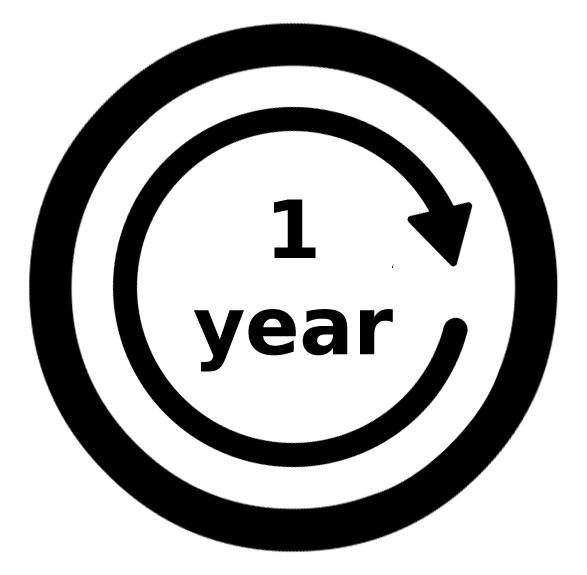}
		\caption{\centering \footnotesize One year conservation}
	\end{subfigure}
	~
	\begin{subfigure}[h]{0.2\textwidth}
		\centering	\includegraphics[scale=.1]{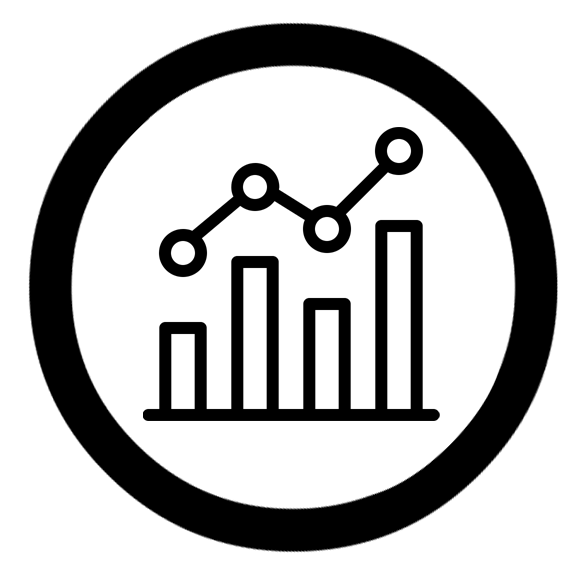}
		\caption{\centering \footnotesize Audience measurement}
	\end{subfigure}
	~
	\caption{Excerpt of the Privacy Tech icons}
	\label{fig:privacytech}
\end{figure}

Many privacy advocates also contributed to this area and provided numerous sets of icons (see~\cite{aaron_privacy_2009, raskin_privacy_2011}).
For instance, Mehldau \cite{mehldau_icons_2007} developed a set of 30 privacy icons describing the type of data, third-parties, the purpose of processing and the retention time.
% (see Figure~\ref{fig:mehldau}). 
Recently, a set of privacy icons was designed by Privacy Tech~\cite{privacy_tech_privacy_2018}. 
This set considers many types of data, as well as advanced representations of sharing, such as adequacy transfer (see Figure \ref{fig:privacytech}).

%\paragraph{Privacy advocacy}
%Many privacy advocates also contributed to this area and provided numerous sets of icons, such as Mehldau \cite{mehldau_icons_2007} who developed a set of 30 privacy icons, describing the type of data, third-parties, the purpose of processing and the retention time (see Figure~\ref{fig:mehldau}). 
%Generally speaking, icons within this subcategory can express more items, and items more related to \nlpp.
%For instance, Aaron \cite{aaron_privacy_2009} came up with a set of icons that can only express three types of data, whether data may be disclosed to third-parties or not (see Figure \ref{fig:aaron}) \etc
%Raskin \cite{raskin_privacy_2011} developed a set of icons for Mozilla. The type of data is not considered, and only the retention time, third-party use, ad networks, and law enforcement are considered (see Figure~\ref{fig:raskin}).
%Recently, a set of privacy icons was designed by Privacy Tech~\cite{privacy_tech_privacy_2018}. 
%This set considers many types of data, as well as advanced representations of sharing, such as adequacy transfer (see Figure \ref{fig:privacytech}). 
%It considers many common items, but not the rights of DS nor policy change --- these items may be seen as less relevant for data transparency, even though mandatory under the GDPR.

%\begin{figure}[!ht]
%	\centering
%	\includegraphics[scale=.45]{Figures/mehldau_icons.pdf}
%	\caption{Mehldau's set of icons}
%	\label{fig:mehldau}
%\end{figure}

%\begin{figure}[!ht]
%
%\end{figure}

%\paragraph{Business sector}
Another notable example of privacy icons are the android permissions~\cite{google_android_nodate}, created by Google. 
They present icons combined with simple natural language.
% (see Figure \ref{fig:android}). 
For each application installed on a mobile phone running Android, the permission manager presents a short graphical policy. 
Only little information is presented (the type of data collected, and processing in recent versions, but not the purpose for instance), and DS have to look into the natural language privacy policy in order to find more information.

\subsubsection*{Standardized notices}
\begin{figure}[!ht]
	%	\hspace{-1cm}
	\centering
		\includegraphics[scale=.6]{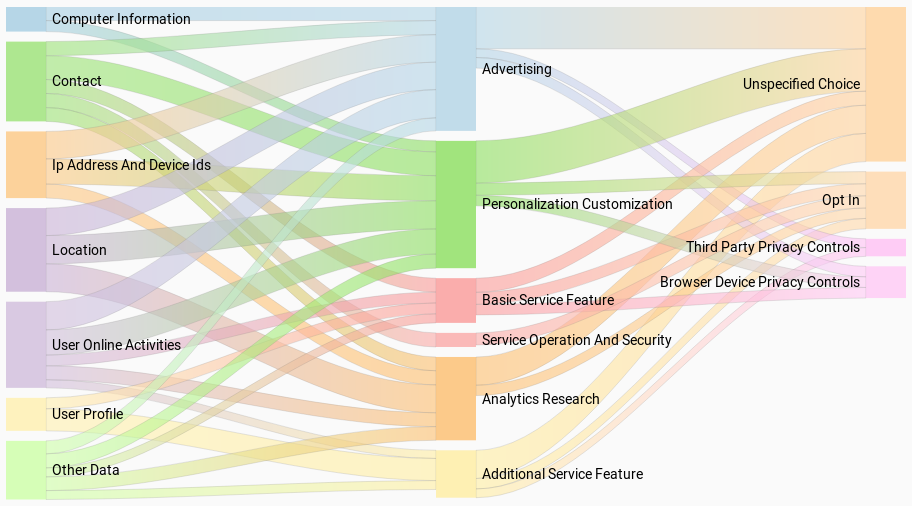}
%		\caption{}
	\caption{Example of a flow diagram in Polisis}
        \label{fig:polisis_flow}
\end{figure}

Another line of work considers the content described by the taxonomy as standardized notices.
These standardized notices are often represented in tables~\cite{kelley_nutrition_2009}, but the key concept is the common vocabulary among notices.
Kelley \etal~\cite{kelley_nutrition_2009} represent policies in a table such as nutrition labels observed on food packaging.
% (see Figure \ref{fig:nutrition}). They developed a privacy nutrition label based on P3P, with the goal of providing efficient and well-organized privacy information. 
They present the fine-grained information in a table such as nutrition labels observed on food packaging. 
Polisis by Harkous \etal~\cite{harkous_polisis_2018} can represent the \nlpp~as a combination of icons,
% (see Figure \ref{fig:polisis_icons}), 
highlights of the corresponding paragraphs in the \nlpp, and a flow diagram (see Figure \ref{fig:polisis_flow}). 
Because Polisis relies on supervised machine-learning, \ie, on a labeled corpus, it classifies \nlpp in a standardized way.
Such summarization has been accomplished by PrivacyCheck as well~\cite{zaeemPrivacyCheckAutomaticSummarization2018}.

%\begin{figure}[!ht]
%	%	\hspace{-1cm}
%	\centering
%	\begin{subfigure}[h]{0.5\textwidth}
%		\centering
%		\includegraphics[scale=.3]{Figures/polisis.png}
%		\caption{Excerpt of Polisis icons}
%		\label{fig:polisis_icons}
%	\end{subfigure}
%	~
%	\begin{subfigure}[h]{0.5\textwidth}
%		\centering
%		\includegraphics[scale=.25]{Figures/polisis2.png}
%		\caption{Example of a flow diagram}
%		\label{fig:polisis_flow}
%	\end{subfigure}
%	~
%	\caption{Polisis}
%\end{figure}

Emami-Naeini~\etal~\cite{emami-naeiniExploringHowPrivacy2019} conduct a survey in order to rank the factors of IoT devices purchase.
They determine that security and privacy were among the most important factors of purchase, and consequently developed an IoT privacy label to improve information visualization.
% (see Figure~\ref{fig:iot_label}).
Cranor analyzes the impact of the development of standardized mechanisms of notice and choice in~\cite{cranor_necessary_2012}, and more specifically the efforts conducted around P3P.
Cranor reconsiders the advances made in standardization,~\footnote{Note that icons are considered as part of standardization efforts in~\cite{cranor_necessary_2012}.} as well as the limitations, lack of adoption and enforcement.

\subsubsection*{Rating solutions}
\label{sec:graph_rating}
Certain \gpp~do not focus on expressing the items in our taxonomy, but present extra information related to privacy policies~\cite{van_den_berg_what_2012, sweeney_sharing_2015, hagan_user-centered_2016}, often a judgment of the risk level associated to a DC policy, or a comparison between DS and DS policies. We denote them \textit{rating solutions}.
%Both academics and privacy advocates contributed to the solutions.

\begin{figure}[!ht]
	\centering
	\begin{subfigure}[h]{0.2\textwidth}
		\includegraphics[scale=1]{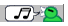}
		\caption{\centering \footnotesize Matching policies}
	\end{subfigure}
	~
	\begin{subfigure}[h]{0.21\textwidth}
		\includegraphics[scale=1]{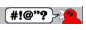}
		\caption{\centering \footnotesize Conflicting policies}
	\end{subfigure}
	~
	\begin{subfigure}[h]{0.2\textwidth}
		\includegraphics[scale=1]{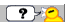}
		\caption{\centering \footnotesize Uncertain decision}
	\end{subfigure}
	~
	\begin{subfigure}[h]{0.2\textwidth}
		\includegraphics[scale=1]{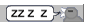}
		\caption{\centering \footnotesize Add-on disabled}
	\end{subfigure}
	~
	\caption{Privacy Bird}
	\label{fig:privacy_bird}
\end{figure}

%\paragraph{Academia}
Privacy Bird~\cite{cylab_usable_privacy_and_security_laboratory_privacy_nodate} is one of the first rating solution. 
It consists of a colored bird, where the color indicates the
matching (green for a match between the DS policy and the website's DC policy, red for conflict, yellow for uncertain, gray when disabled)
(see Figure \ref{fig:privacy_bird}).

It is represented as an add-on for Internet Explorer, restricted to Microsoft Windows. 
A dedicated website~\cite{cylab_usable_privacy_and_security_laboratory_privacy_nodate} provides an explanatory tour as well as a feature named \textit{privacy finder}: Privacy Bird is then used as an indicator when browsing the web~\cite{byersSearchingPrivacyDesign2005,cranor_necessary_2012}.
Privacy finder displays the search results of a search engine, combined with the analysis of Privacy Bird. 
The bird was placed alongside search results, and was influential in the choice of shopping websites~\cite{tsaiEffectOnlinePrivacy2011a}, notably when the items being purchased were likely to have privacy concerns~\cite{egelman_timing_2009}. 
DS could rank the results according to the matching between their DS policy and the websites' DC policies.
%Van den Berg and Van der Hof \cite{van_den_berg_what_2012} devised a wheel whose spokes show how data is handled (see Figure \ref{fig:privacy_wheel}). Their solution highlights fairness of processing rather than transparency: it issues a judgment on the processing, but shows little information with respect to what data is collected, by whom, and for what purpose.
%Sweeney \etal \cite{sweeney_sharing_2015} proposes a simplified interface for access requirement to medical data called Datatags --- further developed in \cite{bar-sinai_datatags_2016} (see Figure \ref{fig:datatags}). To each piece of data can be associated a tag presenting the risks, the security features associated, and the credentials required to access it.
%Hagan describes in \cite{hagan_user-centered_2016} a user-centered privacy policy design project. 
%The proposals include for instance a Visual Data Privacy Diagram to intuitively visualize data flow, Multi-character stories to present concrete situations (see Figure \ref{fig:hagan}), and Context-Specific Alert for a selection of common questions regarding location data. 
%The proposition cannot and does not intend to express items defined in Section \ref{sec:natural_language}, but attempts to increase DS awareness about consequences of data processing.
In~\cite{pilot_apf_2019}, Pardo \& Le M\'etayer present a web
interface to inform DS about the potential risks of their privacy
policies.
The interface is composed of a user-friendly form for DS to input
their privacy policies and a set of risk analysis questions, \eg,
``Can company X collect my data?''.
% (see Figure~\ref{fig:web-app-risk-analysis}).
%
%DS simply need to click on ``Analyze'' to automatically obtain the answer to the questions.
%
Additionally, DS may introduce risk assumptions in order to specify
possible misbehaviors that the collecting parties can perform.
In Section~\ref{sec:machines}, we describe in detail the underlying privacy language and the automatic risk analysis.
%\paragraph{Privacy advocacy}
The ToS;DR initiative helps DS understanding the risks associated to a DC policy~\cite{terms_of_service;_didnt_read_terms_nodate}.
It started in 2011 during the Chaos Communication Camp. 
ToS;DR comprises not only icons, but also results from crowdsource analyses in simple language. 
The idea of the project is to assess the data practices of web services by giving them badges, awarded by the project's community. 
Once a service has enough badges to assess the level of protection of their terms for users, a class is automatically assigned by pondering the average scores.
% (see Figure \ref{fig:tosdr}).

\subsection{Tools}
\label{sec:graph_tools}
Tools for representing \gpp are tailored to the web, and are often found as add-ons for web browsers. 

Privacy Bird is represented as an add-on for Internet Explorer, restricted to Microsoft Windows. 
ToS;DR is also an add-on, for both Firefox and Chrome, as it ranks policies based on crowdsourced analyses by a community directly within the web browser. 
A website has been built to present Polisis~\cite{polisis_polisis_nodate}, and add-ons for Chrome~\cite{polisis_chrome_nodate} and Firefox~\cite{polisis_firefox_nodate} are available (the add-ons redirect to the corresponding part of the website).
%Only one tool is made of icons, unlike the others which represent rating solutions. 
The add-on ''Disconnect Privacy Icons``~\cite{disconnect_privacy_2016}, in collaboration with TRUSTe, which evolved from Raskin's set of icons for Mozilla~\cite{raskin_privacy_2011},
% (see Figure \ref{fig:raskin}), 
provided an interactive and comprehensive view of privacy policies within the browser. The add-on would display icons according to a website privacy policy if the website complies with the solution. 

All of these tools focus on the web, whereas the IoT is left unchallenged in that respect.

%Schaub \etal~\cite{schaub_design_2015} introduced a Design Space for Privacy Notices to foster understanding of \textit{privacy notices}, which can be understood as \gpp~in our sense. They propose guidelines for developing graphical policies.%, and their work can be considered as a tool.
%They distinguish the \textit{timing} (at setup, just in time, context-dependent, periodic, persistent, and on-demand), the \textit{channel} (primary, secondary, and public), the \textit{modality} (visual, auditory, haptic, and machine-readable), and the \textit{control} brought by notices (blocking, non-blocking, and decoupled). They argue that each privacy notice should be thought in these terms, and that an appropriate use combined with user studies would participate in a better understanding from DS.

\subsection{Benefits}
\label{sec:graph_benef}
\Gpp cannot be seen as legal commitments because they lack precise meaning, but they have other benefits: they can foster understanding.

\paragraph*{Designed for lay-user understandability} Many solutions coming from privacy advocates (\eg,~\cite{mehldau_icons_2007, raskin_privacy_2011,aaron_privacy_2009,raskin_making_2010}) aim to provide intelligible information to lay-users.
%: ``In order for privacy policies to have meaning for actual people''
%~\cite{raskin_making_2010}. 
These solutions were built with the will to popularize \nlpp, and were designed to be understood quickly and take simplicity account. It is also the case for academic solutions such as the privacy labels \cite{kelley_nutrition_2009}, which ``allows participants to find information more quickly and accurately''. 
Based on the principle that existing \nlpp~do not convey intelligible information about data collection and processing, Kelley \etal strove to provide a universal solution.
%: ``Our only requirement was that English be the participant's native language''.
\Gpp~can also convey intelligible notices for scientists and physicians using sensitive datasets, such as the DataTags \cite{sweeney_sharing_2015,bar-sinai_datatags_2016}. Their solution includes a simplified interface for access requirement to medical data, as this type of data is mostly restricted to medical practitioners and researchers.

%\subparagraph*{\textbf{Measuring understanding:}}
% \paragraph*{Measuring understandability}
Attempts were made to analyze what icons were recognizable and to measure their reliability. 
Egelman \etal~\cite{egelman_is_2015} crowdsourced privacy indicators for the IoT. In their study, they found out that some icons are well-recognized (\eg, the camera symbol was recognized by more than 95\% of participants as representing \textit{video recording}), while others are not (only 3.6\% recognized the \textit{voice command \& control} icon). 
%The PrimeLife project also proposed a set of privacy icons (\cite[Chapter 15]{camenisch_privacy_2011}), that they tested in order to determine if they were understood. They concluded that clear icons with few details were preferred. 
Kelley \etal~\cite{kelley_standardizing_2010} conducted a user study, to refine their privacy label. They compared the accuracy of information retrieval between their proposition and \nlpp~in natural language. As a result, they purposely combined simple natural language to prevent confusion, notably for the terms \textit{opt-in} and \textit{opt-out}.~\footnote{Note that they also test the speed of retrieval, as well as comparisons between DC policies in addition to information retrieval.} 
%Motti and Caine attempt in~\cite{mottiVisualVocabularyPrivacy2016} to create a visual vocabulary for privacy by examining online images related to privacy.
%They analyze online images produced by users (such as Instagram), UI designers (\eg, Material Design icons) and content producers (Shutterstock for instance).
%They identify five codes, \ie, categories: action, objects, organizations, people, and abstract concepts.
%However, understanding of these icons and the vocabulary they represent is not tested.
A promising attempt to measure understandability has been conducted in~\cite{rossiDaPISOntologyBasedData2019}.
They performed three evaluations of their icon set in order to improve the recognition of icons.
However, they regret the lack of diversity in the participants' panel, notably for the educational level.

\subsection{Limitations}
\label{sec:graph_limit}
\paragraph*{Ambiguity} 
Though accessible to and often designed for lay-users, \gpp may be interpreted in different ways, thus leading to ambiguities (see Section \ref{sec:graph_content}). 
The same icon can be interpreted in different ways according to the differences in culture, education level, or context \etc 
For instance, a euro symbol~\EUR~ can represent the commercial use of collected data, or that DS will be paid for having her data collected. 
Little has been done to produce a reasonably recognized set of icons for privacy --- \textit{e.g.} validated by a user study --- despite the attempts of \cite{egelman_is_2015} and \cite[Chapter 15]{camenisch_privacy_2011} to see what were the most recognizable icons, and of \cite{kelley_standardizing_2010} to provide a graphical policy where results could be found accurately.
The three stages evaluation of Rossi and Palmirani~\cite{rossiDaPISOntologyBasedData2019} however leads the path to less ambiguous \gpp.

\paragraph*{Incompleteness} \Gpp~are limited by their restricted scope. As seen in Section \ref{sec:graph_content}, existing \gpp~are not as expressive as \nlpp, due to the limited number of icons available. Some aspects are rarely mentioned, others only in complementary text and not in the graphical part of the policy.
One aspect in particular is never mentioned in \gpp~(policy change), and another is considered in only one work (legal basis).

\paragraph*{Claim over legal compliance}
Some \gpp, such as cookie consent notices, have been used to claim legal compliance to retrieve consent.
Degeling~\etal~\cite{degelingWeValueYour2019} observed that a significant part (16\%) of websites added cookie consent notices after the GDPR, but these notices do not always comply with transparency requirements according to Utz~\etal~\cite{utzInformedConsentStudying} as they tend to use Dark Patterns~\cite{grayDarkPatternsSide2018} to lure DS into giving their consent.
%\footnote{Dark Patterns are ``instances where designers use their knowledge of human behavior (\eg, psychology) and the desires of end users to implement deceptive functionality that is not in the user’s best interest''~\cite{grayDarkPatternsSide2018}.}

%\paragraph{Solutions}
%The ambiguity of graphical representations is hard to overcome, but it can be reduced by standardization efforts. For instance, a graphical privacy policy answering the requirements of ISO 9186-2:2008 \cite{iso_graphical_2008} for testing perceptual quality will be likely to be understood. Another way to tackle this limitation would be to combine graphical representations with everyday natural language, as \cite{van_den_berg_what_2012} proposed. They argued after a user study that everyday natural language is better understood by DS than icons, and developed a wheel of privacy preferences displaying data practices of web services, combining a visual and a textual approach. They came to that conclusion after a study they conducted, but acknowledged the surprising contrast with previous studies. In practice, numerous \gpp~use a combination of pictures and text (see Table \ref{tab:framework_graphical}), but many solutions do not consider explanations in natural language as a core feature. With regard to their restricted scope, this apparent limitation of \gpp~does not have to be one, as long as they can express all the items required by law, and for intelligibility.\raul{This paragraph should go on insights/perspective.}

%\noindent
%\textbf{Summary.~}

\subsection*{Summary}

\label{subsec:gpp-sum}
Graphical privacy policies are promising for conveying summarized versions of \nlpp, and they can rely on user-friendly tools to be adopted. However, they should come with explanations to ensure human understanding and mitigate their restricted content. 
See Table \ref{tab:framework_graphical} for a visual and global overview of our study on graphical policies. 
%The items of the taxonomy absent of the surveyed work are omitted.
%Not all items of the taxonomy are considered: those not appearing in any of the surveyed works are omitted.

\newcolumntype{R}[2]{%
	>{\adjustbox{angle=#1,lap=\width-(#2)}\bgroup}%
	c%
	<{\egroup}%
}
\begin{table}[!t]
	\normalsize
	\centering
	% \hspace*{-2cm}
        \begin{tabular}{l|llll|ll|llllll|}
		\multicolumn{1}{R{45}{1em}}{} & \multicolumn{1}{R{45}{1em}}{Icons} & \multicolumn{1}{R{45}{1em}}{Simple text} & \multicolumn{1}{R{45}{1em}}{Rating} & \multicolumn{1}{R{45}{1em}}{Standardized notice} & \multicolumn{1}{R{45}{1em}}{DC} & \multicolumn{1}{R{45}{1em}}{DS} & \multicolumn{1}{R{45}{1em}}{$1^{st}$ party} & \multicolumn{1}{R{45}{1em}}{$3^{rd}$ party} & \multicolumn{1}{R{45}{1em}}{DS rights} & \multicolumn{1}{R{45}{1em}}{Data retention} & \multicolumn{1}{R{45}{1em}}{Data security} &
		\multicolumn{1}{R{45}{1em}}{Legal basis} \\ \midrule %\hline
		\multicolumn{1}{l|}{Privacy Bird~\cite{cylab_usable_privacy_and_security_laboratory_privacy_nodate}} & \checkmark &  & \checkmark & & \checkmark & \checkmark & \textemdash & \textemdash & \textemdash & \textemdash & \textemdash & \textemdash \\ %\hline
		
		\rowcolor{Gray}
		\multicolumn{1}{l|}{Rundle \cite{rundle_international_2006}} & \checkmark & \checkmark & & &  & \checkmark & \LEFTcircle & \Circle & \CIRCLE & \Circle & \LEFTcircle  & \Circle \\ %\hline
		
		\multicolumn{1}{l|}{Mehldau \cite{mehldau_icons_2007}} & \checkmark &  & & & \checkmark &  & \CIRCLE & \LEFTcircle & \Circle & \CIRCLE & \CIRCLE  & \Circle \\ %\hline
		
		\rowcolor{Gray}
		\multicolumn{1}{l|}{Privacy Commons \cite{aaron_privacy_2009}} & \checkmark & & &  & \checkmark &  & \LEFTcircle & \LEFTcircle & \Circle & \Circle & \Circle  & \Circle \\ %\hline
		
		\multicolumn{1}{l|}{Privacy Nutrition Label \cite{kelley_nutrition_2009}} & & \checkmark & & \checkmark & \checkmark &  & \CIRCLE & \LEFTcircle & \Circle & \Circle & \Circle  & \Circle\\ %\hline
		
		\rowcolor{Gray}
		\multicolumn{1}{l|}{Primelife \cite[Chapter 15]{camenisch_privacy_2011}} & \checkmark &  & & & \checkmark &  & \textemdash & \textemdash & \textemdash & \textemdash & \textemdash  & \Circle \\ %\hline
		
		\multicolumn{1}{l|}{Raskin \cite{raskin_privacy_2011}} & \checkmark &  & & & \checkmark &  & \LEFTcircle & \LEFTcircle & \Circle & \CIRCLE & \Circle  & \Circle \\ %\hline
		
		\rowcolor{Gray}
		\multicolumn{1}{l|}{Privicons \cite{konig_privacy_2012}} & \checkmark &  &  & & & \checkmark & \Circle & \LEFTcircle & \Circle & \Circle & \Circle  & \Circle \\ %\hline
		
		\multicolumn{1}{l|}{Privacy wheel \cite{van_den_berg_what_2012}} &  & \checkmark & \checkmark & & \checkmark &  & \Circle & \LEFTcircle & \Circle & \Circle & \LEFTcircle  & \Circle \\ %\hline
		
		\rowcolor{Gray}
		\multicolumn{1}{l|}{Android permissions~\cite{google_android_nodate}} & \checkmark & \checkmark & & & \checkmark &  & \LEFTcircle & \Circle & \Circle & \Circle & \Circle  & \Circle \\ %\hline
		
		\multicolumn{1}{l|}{``Is this thing on?''\cite{egelman_is_2015}} & \checkmark &  & & & \checkmark &  & \LEFTcircle & \Circle & \Circle & \Circle & \Circle  & \Circle \\ %\hline
		
		\rowcolor{Gray}
		\multicolumn{1}{l|}{Datatags \cite{bar-sinai_datatags_2016}} &  & \checkmark & \checkmark & & \checkmark &  & \Circle & \Circle & \Circle & \Circle & \CIRCLE  & \Circle \\ %\hline
		
		\multicolumn{1}{l|}{Hagan \cite{hagan_user-centered_2016}} &  & \checkmark & \checkmark & & \checkmark &  & \LEFTcircle & \LEFTcircle & \LEFTcircle & \LEFTcircle & \Circle  & \Circle \\ %\hline
		
		\rowcolor{Gray}
		\multicolumn{1}{l|}{Polisis \cite{harkous_polisis_2018}} & \checkmark & \checkmark & & \checkmark & \checkmark &  & $\CIRCLE_a$ & $\CIRCLE_a$ & $\CIRCLE_a$ & $\CIRCLE_a$ & $\CIRCLE_a$  & \Circle \\ %\hline
		
		\multicolumn{1}{l|}{Privacy Tech~\cite{privacy_tech_privacy_2018}} & \checkmark &  & & & \checkmark &  & \CIRCLE & \LEFTcircle & \Circle & \CIRCLE & \LEFTcircle  & \Circle \\ %\hline
		
		\rowcolor{Gray}
		\multicolumn{1}{l|}{DaPIS~\cite{rossiDaPISOntologyBasedData2019}} & \checkmark &  & & & \checkmark &  & \LEFTcircle & \LEFTcircle & \CIRCLE & \Circle & \Circle  & \CIRCLE \\ %\hline
		
		\multicolumn{1}{l|}{IoT label \cite{emami-naeiniExploringHowPrivacy2019}} & & \checkmark & & \checkmark & \checkmark &  & \CIRCLE & \CIRCLE & \Circle & \CIRCLE & \CIRCLE  & \Circle \\ %\hline
		
		\rowcolor{Gray}
		\multicolumn{1}{l|}{ToS DR~\cite{terms_of_service;_didnt_read_terms_nodate}} &  & \checkmark & \checkmark & & \checkmark &  & $\CIRCLE_a$ & $\CIRCLE_a$ & $\CIRCLE_a$ & $\LEFTcircle_a$ & $\LEFTcircle_a$  & \Circle\\ %\hline
		
		& \multicolumn{4}{c|}{Features} & \multicolumn{2}{c|}{Type} & \multicolumn{6}{c|}{Content} \\ \bottomrule %\cline{2-13} 
	\end{tabular}

	\captionsetup{singlelinecheck=off}
%	\caption[foo bar]{Categorization of \gpp \\ 
%		We use the subscript $_a$ to denote that a solution extensively uses natural language besides graphical representations. \\ 
%%		The subscript $_b$ denotes that Privacy Bird uses icons, although not to specifically express items presented in the taxonomy, as opposed to other solutions within that category (see Section~\ref{sec:graph-icons}).
%		\begin{description}
%			\item[Features] Whether the solution is made of icons or provides ratings about policies, and whether it provides explanation in simple natural language
%			\item[Type](Type of policy) Whether the solution expresses a DC or a DS policy
%			\item[Content] Whether the solution can express the different items enumerated in Section \ref{sec:natural_language}. 
%				We use \CIRCLE\ to denote that the solution can express most or all values;
%				\LEFTcircle\ to denote that the solution expresses few values of the items, and is mostly insufficient;
%				\Circle\ to denote that  the solution cannot express the item; and
%				``\textemdash'' to denote that the material does not permit judging whether the solution can express this item or not.
%                        Note some items of the taxonomy are omitted since no solution includes them.
%		\end{description}
%	}
	\caption[foo bar]{Categorization of \gpp.
	We use the subscript $_a$ to denote that a solution extensively uses natural language besides graphical representations. 
	%		The subscript $_b$ denotes that Privacy Bird uses icons, although not to specifically express items presented in the taxonomy, as opposed to other solutions within that category (see Section~\ref{sec:graph-icons}).
	The column indicates:
	\emph{Features,} whether the solution is made of icons or provides ratings about policies, and whether it provides explanation in simple natural language;
	\emph{Type (Type of policy),} whether the solution expresses a DC or a DS policy;
	\emph{Content,} whether the solution can express the different items enumerated in Section \ref{sec:natural_language}. 
	We use \CIRCLE\ to denote that the solution can express most or all values;
	\LEFTcircle\ to denote that the solution expresses few values of the items, and is mostly insufficient;
	\Circle\ to denote that  the solution cannot express the item; and
	``\textemdash'' to denote that the material does not permit judging whether the solution can express this item or not.
	Note that some items of the taxonomy are omitted since no solution includes them.
%		\vspace*{-1cm}
	}
	\label{tab:framework_graphical}
\end{table}

%%% Local Variables:
%%% mode: latex
%%% TeX-master: "../csur"
%%% End:
\section{Machine-readable privacy policies}
\label{sec:machines}

%% Introduce the content of the section.
%
Many efforts have been devoted to the expression of
\emph{machine-readable privacy policies} --- \textit{i.e.}, privacy
policies that can be automatically processed by computers.
Most of these efforts were made by academics, and result in what has
been called \textit{privacy languages}.
According to Kasem \etal~\cite{kasem-madani_security_2015}, a privacy
language is ``a set of syntax and semantics that is used to express
policies''.
Many privacy languages have been proposed in the past twenty years~\cite{kasem-madani_security_2015,van_de_ven_qualitative_2016}.
We review here the different ways in which privacy languages are used to express machine-readable privacy policies.
In particular, we categorize each work based on:
\begin{inparaenum}[i)]
\item the elements in the taxonomy presented in Section~\ref{sec:natural_language} that it captures;
\item the type of enforcement mechanism it uses and whether it has been implemented;
\item additional tools for policy analysis or comparison;
\item the intended audience of the language, DS or DC; and
\item whether it is intended to be used by lay-users.
\end{inparaenum}
Table~\ref{tab:framework_languages} summarizes our study.

\subsection{Content}\label{subsec:mr-content}

%%% Intro to the section
%
Here we describe the content that machine-readable privacy policies include.
This content is determined by the syntax of the privacy language.
Many languages are defined using machine-readable formats such as XML
so that they can be automatically processed by machines.
Other languages, however, are based on mathematical definitions
(\textit{e.g.}, logical languages), thus enabling the possibility of
reasoning about them --- these languages can easily be expressed in
machine-readable formats due to the lack of ambiguity.
Another important factor is the target audience of a language,
\textit{i.e.}, DC, DS or both.
%
% Some languages are focused on providing a platform for DS and DC to
% express privacy policies, whereas others are desinged to model privacy
% regulations such as HIPAA or COPPAA.
%
In what follows, we describe the content of machine-readable languages
(according to the items defined in
Section~\ref{sec:natural_language}), the format used to express the
policies and their target audience.

%%%%%%%%%%%%%%%%%%%%%%% related work from APF paper %%%%%%%%%%%%%%%%%%%%%%%%%
%% Access control languages
%
Access control languages such as XACML~\cite{anderson_extensible_2003}
and RBAC~\cite{SCFYrbacm96} have been among the first languages used
for the specification of machine-readable privacy policies.
Typically, these policies include the datatype to which they apply,
and the set of entities with access privileges.
Some extensions such as GeoXACML~\cite{geoaxcml} include conditions
depending on geolocation information.
%
%, \eg, ``Alice can only access data from Lyon''.
%
% However, none of these languages captures concepts such as retention
% time, purpose or transfers that are in the privacy policy taxonomy 
% described in Section~\ref{sec:natural_language}.
%
% In other words, access control languages cannot impose any usage
% constrains after data has been accessed.
%
%%% Usage control
Usage control (UCON)~\cite{PSuucm04,PHBduc06} extends access control
with 
% so that it is possible to express how the data may be used after being
% accessed.
%
% UCON introduces
the notion of \emph{obligations}, \textit{i.e.},
actions to be executed after data has been received --- \textit{e.g.},
``do not transfer data item $i$ to Service X'' or ``remove data on
25/05/2018''.
These obligations make it possible to express items such as retention
time, purpose and allowed data transfers.
The Obligation Specification Language (OSL)~\cite{HPBSWplduc07} is a
fully-fledged UCON language.
OSL leverages Digital Right Management systems
(DRMs)~\cite{drm_privacy} to enforce the obligations in UCON policies.
%
%%% Neither access control nor usage control where meant to be used to
%%% express privacy policies
% Neither access control nor UCON were developed with the idea of
% expressing privacy policies in mind.
%
% For instance, these languages do not offer mechanisms to describe DS policies.
%
% They are mostly used by DC to define their policies.
%
% New policy languages focused on expressing privacy policies appeared to address this problem.
%
% See column \emph{Audience} in Table~\ref{tab:framework_languages} for
% the target audience of each of the surveyed languages.\victor{Why is this sentence here?}
Both access control and UCON are used by DC to define their policies,
and do not offer mechanisms for DS to express their policies.

%%% Privacy languages
Several languages focused on expressing privacy policies have been
developed.
One of the first is the ``Platform for Privacy Preferences''
(P3P)~\cite{cranor_platform_2002}.
P3P appeared as a policy language for the web.
%
% It allows clients to declare their privacy preferences, and online
% service providers (mostly websites) to inform how they use customers' data.
%
P3P policies are specified in XML format, and include: purpose,
retention time, and \emph{conditions}.
Conditions may be opt-in and/or opt-out choices for DS, or
preferences based on enterprise data --- \textit{e.g.}, DS's
credit or service usage.
Many extensions to P3P have been proposed~\cite{langheinrich2002appel,
  ashley_e-p3p_2002, agrawal_xpath-based_2003} 
% , where its syntax has been extended
--- for instance, E-P3P~\cite{ashley_e-p3p_2002} extends P3P with
obligations \emph{\`{a} la} UCON.
After P3P, new languages with similar syntax have been proposed:
``Enterprise Policy Authorization Language''
(EPAL)~\cite{ashley_enterprise_2003}, ``An Accountability Policy
Language'' (A-PPL)~\cite{appl}, ``Customer Profile Exchange''
(CPExchange)~\cite{bohrer_customer_2000}, ``Privacy Rights Markup
Language'' (PRML)~\cite{zero-knowledge_privacy_2001},
``Purpose-to-Use'' (P2U)~\cite{iyilade_p2u_2014} and ``Layered Privacy
Language'' (LPL)~\cite{LPLgbkb18}.
These languages are similar than P3P in terms of content, but bring
numerous enhancements in terms of usability and enforcement
(see Section~\ref{subsec:mr-tools}).
Formal privacy languages (\emph{formal languages} in the following)
comprise a different approach to express privacy policies.
CI~\cite{nissenbaum_privacy_2004}, PrivacyAPIs~\cite{MGLpaactavlpp06},
SIMPL~\cite{le_metayer_formal_2008}, PrivacyLFP~\cite{DGJKDelshgpl10},
S4P~\cite{becker_s4p_2010}, QPDL~\cite{van_de_ven_qualitative_2016},
and PILOT~\cite{pilot_apf_2019} are languages that have their syntax
and semantics defined mathematically.  More precisely, they use formal
languages such as \emph{Linear Temporal Logic}~\cite{ltl_first_order},
\emph{First-Order Logic}~\cite{ltl_first_order} or \emph{Authorization
  Logic}~\cite{authorization_logic}.
However, not all of these formal languages have the same focus.
S4P, SIMPL and PILOT are focused on expressing DS and DC policies.
%
% Thus, they do not differ much in content from the languages mentioned
% in the paragraphs above.
%
Similarly to the languages above, it is possible to express types of
data, conditions, purpose, retention time and allowed data transfers.
Conditions are often more sophisticated than that of the previous
languages as they are based on logical languages.
%
% For instance, PILOT makes it possible to include spatio-temporal
% conditions which allow DS and DC to describe when, where, and by which
% devices data may be collected.
%
On the other hand, CI, PrivacyAPIs and PrivacyLFP focus on encoding
privacy regulations such as
HIPAA~\cite{united_states_congress_health_1996},
COPPA~\cite{federal_trade_commission_childrens_2013} or
GLBA~\cite{united_states_congress_grammleachbliley_1999}.
As a consequence, their expressive power is greater than languages
focusing on DS and DC policies.
They include temporal operators that make it possible to express policies
about past and future events.
For example, Barth \etal~\cite{barth_privacy_2006} express the
following statement from COPPA ``[...] an infant can only send
identifiable information to a website, if her parent have previously
sent their consent for data collection''.
QPDL is a meta-language to reason about privacy languages.
While privacy policies can be expressed in QPDL, it is not its intended use --- it was designed to formally reason about policy languages.
%
% The language was conceived as a framework to formally reason about different policy languages.
%
% , \ie, to compare the expressive power of different languages.
%
% The goal of these formal semantics is to be able to prove global
% correctness properties such as ``\dcs always use \ds policies
% according to their policies''.
%
% While this semantics is well-suited for its intended purpose, they
% cannot be directly used to develop policy enforcement mechanisms.  In
% contrast, we provide a \promela model in
% Section~\ref{sec:risk-analysis}---capturing the execution model of
% \pilot (cf. Section~\ref{sec:model})---that can be used as a reference
% to implement a system for the enforcement for \pilot policies.  In
% addition, the above abstract languages, which were proposed before the
% adoption of the GDPR, were not conceived with its requirements in
% mind.
%
%%% Programming languages with limited support for privacy policies.
Jeeves~\cite{yang_language_2012} is a programming language with
built-in support for a limited form of privacy policies.
It allows programmers to specify confidentiality conditions based on
the execution context.
For instance, in a double-blind conference management system, only organizers can see paper authors until the review process is completed.

\subsection{Tools}\label{subsec:mr-tools}
%Most means of machine-readable policies expression are mathematical structures, or protocol specifications. We will focus on the underlying mathematical structure of privacy languages. The mathematical structure (hereafter logic) of privacy languages can be derived from their syntax, and semantics if they possess one: what can be expressed and what is the precise meaning of what is expressed. However, a privacy language without semantics — formal or not — is not necessarily without meaning, but could be interpreted in different ways. It is possible to know its scope if a language has a syntax, and it is even possible to prove assumptions if it possesses formal semantics. Proving properties of a privacy language often amounts to formally demonstrate that only required data will be collected (and processed if needs be), and so for any type of data.

In this section, we describe the mechanisms used to enforce
machine-readable privacy policies, and existing tools to compare and perform analyses on policies for instance.

\paragraph*{Formal Semantics}
Formal languages give meaning to their privacy policies by means of
\emph{formal semantics}.
Typically, these semantics define what events may be executed
depending on the privacy policies selected by the actors interacting in
the system. %, \eg, DS and DC.
%
% There are several ways to express semantics formally.
%
%For instance,
SIMPL, PILOT, S4P and CI use trace semantics, \textit{i.e.}, they
define the sequences of events (traces) that respect the privacy
policies of DS and DC.
%
% PILOT uses small step operational semantics that define what events
% may be executed given the state of the system and privacy policies.
%
Jeeves has its semantics formalized using lambda
calculus~\cite{lambda_calculus}.
Rei semantics is defined in Prolog~\cite{prolog}.
Though precise and unambiguous, 
% formal semantics are not directly executable in most cases.
there is a gap between the definition of formal semantics and the real implementation.
Nevertheless, this gap may be very small, \textit{e.g.}, Jeeves lambda
calculus semantics were implemented as a Scala library, Rei's
semantics are encoded in Prolog, and PILOT semantics are implemented
as a Promela model~\cite{promela}.

\paragraph*{Informal Semantics}
Access control, UCON and privacy dedicated languages have their
enforcement mechanisms specified as W3C specifications, specification
languages such as UML, or they are simply implemented using a general
purpose programming language.
All these languages %have in common that they 
use \emph{request
  evaluation engines} to enforce privacy policies.
%
% As opposed to the languages described in the paragraph above, these
% engines do not feature formal semantics.
%
Request evaluation engines take a \emph{data request} and evaluate
whether the requester may access the data based on the privacy
policies.
The content of data requests depends on the language.
For instance, in RBAC, data requests contain type of data and the role
of the requester.
If the role of the requester matches one of the roles allowed by the
policy associated with the data, then data can be accessed.
%
% Usually, data requests include more information, \eg, P3P data
% requests include data type, purpose of usage, requesting user, and the
% action to be performed (\eg, read, write, delete, \etc).
%
Most languages do not have mechanisms to enforce that data will be
used according to the policies after being collected --- \textit{e.g.},
checking that data is deleted before the retention time
% , or used for the specified purposes
--- but there are some exceptions.
LPL erases automatically data from the central repository after the
retention time has elapsed.
UCON-based languages
%, such as OSL,
use DRM to enforce obligations.
%
%An important benefit of these languages is that they are implementedand ready to be deployed.
%
% A common factor of all these languages is that their request
% evaluation engines have been implemented and are ready to be deployed.

\paragraph*{Policy comparison}
For some languages, algorithms have been devised to automatically compare policies.
The goal is to determine, given two policies, which one is more restrictive.
For example, a policy that allows data processing for research
purposes during 7 days is more restrictive than a policy that allows
data processing for advertisement and research during 90 days.
Policy comparison is necessary to mechanize consent management.
If a DC policy is more restrictive than a DS policy, then DS privacy preferences are satisfied.
%
% This step, although insufficient, is necessary for consent to be legally valid.
%
% Examples of such languages include
EPAL, P3P and PILOT include algorithms and tools to compare policies.
%
% In fact, the graphical tool Privacy Bird (see Section~\ref{sec:graph_tools}) uses P3P's comparison algorithm to provide visual feedback to DS.
%
CI, SIMPL and S4P follow a different approach.
They define how restrictive a policy is, based on its
semantics.~\footnote{Using trace semantics it is possible to compare
  policies based on the set of traces satisfying the policy.
  The less traces a policy satisfies, the more restrictive it
  is.}
Languages that do not distinguish DS and DC policies --- such as RBAC,
A-PPL, or OSL --- do not have algorithms to compare policies.
This is not surprising, their goal is to enforce a policy typically
defined by DC or system administrators.

\paragraph*{Analysis tools}
Formal languages often come with tools to perform different types of
automatic analyses.
PILOT uses model-checking~\cite{model_checking} to perform risk
analysis.
Given a DS policy and a set of risk assumptions, such as ``Company X
may transfer data to Company Y'', it is possible to automatically
answer questions such as ``Can Company Z use my data for
advertisement?''.
% or ``Can my data be collected by Company Z?''.
%
Rei comes with a Prolog interface where queries such as the above can
be asked.
PrivacyAPIs also uses model-checking to automatically verify
properties about the privacy regulation HIPAA.
It can, for instance, be used to determine who can access patients
medical files depending on their content or role.

% Most means of machine-readable policies expression are mathematical structures, or protocol specifications. However, a few privacy languages come with complementary tools. These tools can be implementations in programming languages, but it is also possible to find an add-on. As a matter of fact, P3P was extended with the web browser plug-in Privacy Bird, already mentioned in Section~\ref{sec:graph-tools}. \\

% Implementations in programming languages are also named model checkers. \victor{Definition of model checkers} For instance, Rei by \citet{kagal_policy_2003} comes with a Prolog interface, and the policy engine has a Java wrapper. It makes this language more concrete as it can actually be implemented. Another language, Jeeves, designed by \citet{yang_language_2012}, comes with a Scala embedding. Jeeves was implemented as a Scala library to test implementations. \\

% In general, means of expressing machine-readable privacy policies lack proper deployment tools, and this lack of effective implementation is strongly correlated to their lack of adoption. Column \textit{other} of Table \ref{tab:framework_languages} summarizes the tools.

\subsection{Benefits}
\label{subsec:mr-benefits}
Machine-readable privacy policies have four main benefits:
\begin{inparaenum}[1)]
\item they can be automatically enforced;
\item they can be audited;
\item it is possible to reason about their correctness; and
\item they make it possible to automate certain procedures.
\end{inparaenum}
In what follows we explain each of these benefits in detail.
%\victor[inline]{@RAUL add that they can give rise to useful analysis tools, notably auditing tools for auditors and DPAs (required if we want to bring up section 5)}

\paragraph*{Enforcement}
As opposed to natural language or graphical policies, machine-readable
policies can be automatically enforced.
As described in Section~\ref{subsec:mr-tools}, all policy languages
have the means to guarantee that data is accessed according to the
policies.
%
% Languages based on UCON or formal languages often provide stronger
% guarantees as they define how data are processed by all the parties
% after data collection.
%
% For example, they ensure that data is only used for purposes in the
% policies or that data is only transferred to allowed entities.
%
% Languages based on request evaluation offer weaker guarantees as they
% only protect access to the data, but not how the receiving party must
% process the data --- only UCON offers limited support via DRMs.
%
Languages based on request evaluation are easier to implement
(cf~\Cref{subsec:mr-tools}), and are consequently more widespread.
%
% Nevertheless, due to their simplicity and ease of implementation,
% request evaluation languages are more widespread.
%
Typically, every party holding personal data must implement a part
of the request evaluation engine.
In general, the implementation of formal languages is more
complicated.
They require tracking actions applied to the data, or inferring what
are the purposes for which data is used --- as opposed to simply
control access to data.

\paragraph*{Auditability}
\Mrpp make it possible to audit whether data is being handled
according to the privacy policies.
This functionality is of great value for DPAs.
Auditing mechanisms are typically implemented as logs recording the
operations performed on sensitive data.
For instance, EPAL requires to create an audit trail of access to keep
track of whom has accessed personal data.
In A-PPL, on the other hand, it is possible to specify \emph{auditable
  operations} such as read or delete, and the enforcement records in a
log every time that such operations occur.
%
%%% Interesting sentence but unrelated to the discussion
% Ensuring the integrity of the logs is an orthogonal issue which is
% crucial for the legal validity of the auditing
% mechanism~\cite{bella_enforcing_2011,schneier_cryptographic_nodate}.

\paragraph*{Correctness}
The lack of ambiguity in policy languages makes it possible to
precisely reason about their correctness, \ie, that data is handled as
stated in the privacy policies.
This is specially true for formal languages.
Their semantics can be used to formally prove certain correctness properties.
For example, S4P, SIMPL and PILOT have been used to prove global
properties such as ``data is never used after its retention time'',
or, ``data is always used according to DS policies''.
Moreover, languages focused on modeling privacy regulation --- CI,
PrivacyAPIs and PrivacyLFP --- can be used to find inconsistencies in
the regulation (if any).
For example, it was possible using PrivacyAPIs to find unexpected
ambiguities in HIPPAA.\footnote{These ambiguities were also found by
  commenters four years after it was enacted~\cite{MGLpaactavlpp06}.}
It is important to remark that there exists a gap between the formal
semantics and its implementation --- technical details not modeled in
the semantics may lead to unforeseen violation of the properties.
Therefore, formal languages should include auditing mechanisms, as the
languages mentioned in the previous paragraph.

\paragraph*{Automation}
\Mrpp allow for automating procedures such as
policy communication and consent management.
Automatic policy communication facilitates transparency by boosting DS
awareness about how their data is being handled --- notably in
ubiquitous systems where passive data collection is the
norm~\cite{das_personalized_2018}.
%
% For instance, Das \etal  propose~\cite{das_personalized_2018}
% Personalized Privacy Assistants for the IoT.
%
% These assistants can inform DS of surrounding IoT devices thanks to
% the machine-readability of the information communicated.
%
Automatic consent management can empower DS if managed in a protective way --- \eg, by mitigating the burden of choice~\cite{sunstein_choosing_2014} --- and facilitate the retrieving of an informed consent for DC.
Cunche \etal~\cite{trustcom_2019} devise a generic framework to manage
informed consent in the IoT, using DS and DC policies based on
PILOT~\cite{pilot_apf_2019}.
Automatic communication of privacy policies also makes possible a
negotiation of privacy choices: DC and DS can interact more quickly by
means of machines.

\subsection{Limitations}\label{subsec:mr-limitations}

The main limitations of machine-readable privacy policies are their
lack of adoption and usability.
As adoption relies among other things on human-understandability,
understandable and usable policies seems to be a condition
\textit{sine qua non} for their adoption.
% and efficiency.

\paragraph*{Human understandability}
One of the most recurring criticism of \mrpp is their lack of human
understandability.
Only a handful of languages such as SIMPL, LPL or PILOT take it into
account: they include a natural language version of each policy.
It is however questionable whether they can actually be understood.
%
% As an example, XPref by \citet{agrawal_xpath-based_2003} aims to
% simplify APPEL recommendations, but still provides an XML scheme to
% express privacy preferences.
%
To put things into perspective, the
OECD~\cite{organisation_for_economic_co-operation_and_development_skills_2016}
conducted a study showing that two third of adults from developed
countries cannot conduct a medium-difficulty task related to ICT
environments.
Although privacy management was not mentioned in the study, it is a
medium-difficulty task, and solutions tackling privacy management must
consider information-illiteracy.
Machine-readable privacy policies should be expressed in languages close
to natural language in order to be understood, or be complemented by
friendly interfaces.
% \victor{maybe not its place} \raul{I think it is ok. It is just part of the discussion on limitations.}
%
% Table \ref{tab:framework_languages} highlights the languages which address this issue in the column \textit{usability}.

%The human comprehension, although a focus in the last years, has to be rethought, because of its implications for user's consent. It cannot be reduced to the comprehension by experts, but needs to be universal. The privacy rulesets specified by \citet{w3_privacy_2009}, presents a machine-readable set of rules for users' privacy, where the categories (called privacy elements) and their components (attributes) are defined as :
%\begin{description}
%	\item[\textbf{Sharing}] internal, affiliates (organizations related to the data controller), unrelated-companies, public
%	\item[\textbf{Secondary use}] contextual, customization, marketing-or-profiling
%	\item[\textbf{Retention}] no (actually for retained for a baseline period \textit{e.g.} 35 days), short (limited time), long (unspecified or indefinite amount of time)
%\end{description}
%
%The combination of these attributes can give a desired policy for a certain type of data, for example defining a profile-based advertising use of data would be represented as :
%\begin{itemize}
%	\item sharing=internal
%	\item secondary-use=marketing-or-profiling
%	\item retention=long
%\end{itemize}

\paragraph*{Lack of adoption}
Another pitfall for machine-readable privacy policies is their lack of
adoption.
It is arguably a consequence of poor human understandability.
%
% As mentioned in Section \ref{sec:machine_stake_adoption}, P3P (and its
% extensions) is the only language that can claim a certain
% recognition.
% Implementation does not automatically mean large scale adoption.
%
Most of the work done on privacy languages had few or no impact, apart
from P3P.
With the other solutions stemming from it (APPEL, E-P3P, \ldots) and
the extension Privacy Bird for Internet Explorer, P3P obtained
recognition out of the academic scope. 
%P3P can claim to have had an impact on the civil society, albeit minor.
%
It has been an official set of specifications of the W3C supported by
the web browser Internet Explorer.
%\footnote{The Electronic Privacy Information Center (EPIC) wrote a report about it, even though it was to highlights its defects.}
%
Note that other languages were published as specifications by companies~\cite{bohrer_customer_2000,anderson_extensible_2003} and can therefore be considered as having had some recognition.
% , even though impact recognition is hard to judge afterwards.\victor{not clear}
%
On the other hand, most formal languages lack a practical and scalable
implementation, making it difficult to use them in practice.
Usability, implementation and widespread recognition are a
rare combination in privacy languages.
%
%The column \textit{adoption} in Table \ref{tab:framework_languages}
%provides a summary of the last paragraph.
%
% But this recognition came thanks to the support of the FTC and the
% W3C.
%
% Adoption is a key issue for Transparency Enhancing Technologies
% (TETs), and solutions aiming at solving an issue should consider the
% \textit{actual} resolution of the problems.

% The main limitations of machine-readable privacy policies are their lack of usability and adoption. As adoption relies among other things on human-understandability, providing understandable and usable policies — in addition with their machine-readability — seems to be a condition \textit{sine qua non} for their adoption and efficiency.

%\noindent
%\textbf{Summary.~}
\subsection*{Summary}
\label{subsec:mr-summary}
\Mrpp~can provide means to express unambiguous privacy policies, and
can be enforced as well as audited by authorities.
However, they are often unintelligible for lay-users, which results in
a lack of adoption.
We provide a visual and global overview of machine-readable policies
in Table \ref{tab:framework_languages}.
%
% Not all items of the taxonomy are considered: those not appearing in
% any of the surveyed works are omitted.

\newcolumntype{R}[2]{%
	>{\adjustbox{angle=#1,lap=\width-(#2)}\bgroup}%
	c%
	<{\egroup}%
}
\newcommand*\rot{\multicolumn{1}{R{45}{1em}}}% no optional argument here, please!

\begin{table}[!ht]
  \centering
  \scalebox{.84}{
	\begin{tabular}{l|lllll|ll|ll|lllll|}
		\multicolumn{1}{c}{}                                           & \multicolumn{1}{R{45}{1em}}{Usability} & \multicolumn{1}{R{45}{1em}}{Syntax} & \multicolumn{1}{R{45}{1em}}{Enforcement} & \multicolumn{1}{R{45}{1em}}{Implemented} & \multicolumn{1}{R{45}{1em}}{Tools} & \multicolumn{1}{R{45}{1em}}{DS} & \multicolumn{1}{R{45}{1em}}{DC} & \multicolumn{1}{R{45}{1em}}{Time} & \multicolumn{1}{R{45}{1em}}{Space} & \multicolumn{1}{R{45}{1em}}{$1^{st}$ party} & \multicolumn{1}{R{45}{1em}}{$3^{rd}$ party} & \multicolumn{1}{R{45}{1em}}{DS rights} & \multicolumn{1}{R{45}{1em}}{Data security} & \multicolumn{1}{R{45}{1em}}{Data Retention} \\ %\hline
		\midrule
		\multicolumn{1}{l|}{P3P \citep{cranor_platform_2002}}         &                                & XML                         & Informal                       & \yes                             & Comparison                 & \yes                    & \yes                    & \prtl                     & \absn                      & \cmplt                              & \absn                             & \absn                          & \absn                              & \prtl                               \\ %\hline
		\rowcolor{Gray}
		\multicolumn{1}{l|}{CPExchange \citep{bohrer_customer_2000}}  &                                & XML                         & Informal                       &                                  &                            &                         & \yes                    & \absn                     & \absn                      & \cmplt                              & \absn                             & \absn                          & \prtl                              & \cmplt                              \\ %\hline
		\multicolumn{1}{l|}{PRML \citep{zero-knowledge_privacy_2001}} & \yes                           & XML                         & Informal                       &                                  &                            &                         & \yes                    & \absn                     & \absn                      & \cmplt                              & \prtl                             & \absn                          & \prtl                              & \prtl                               \\ %\hline
		\rowcolor{Gray}
		\multicolumn{1}{l|}{APPEL \cite{langheinrich2002appel}}       & \yes                           & XML                         & Informal                       & \yes                             &                            & \yes                    & \yes                    & \prtl                     & \absn                      & \cmplt                              & \absn                             & \absn                          & \absn                              & \prtl                               \\ %\hline
		\multicolumn{1}{l|}{E-P3P \citep{ashley_e-p3p_2002}}          &                                & XML                         & Formal                         &                                  &                            & \yes                    & \yes                    & \prtl                     & \absn                      & \prtl                               & \prtl                             & \absn                          & \prtl                              & \prtl                               \\ %\hline
		\rowcolor{Gray}
		\multicolumn{1}{l|}{Rei \citep{kagal_rei_2002}}               &                                & Formal                      & Formal                         &                                  & Analysis                   &                         & \yes                    & \prtl                     & \prtl                      & \prtl                               & \prtl                             & \absn                          & \absn                              & \prtl                               \\ %\hline
		\multicolumn{1}{l|}{Xpref \citep{agrawal_xpath-based_2003}}   & \yes                           & XML                         & Informal                       & \yes                             &                            & \yes                    &                         & \prtl                     & \absn                      & \cmplt                              & \absn                             & \absn                          & \absn                              & \prtl                               \\ %\hline
		\rowcolor{Gray}
		\multicolumn{1}{l|}{XACML \citep{anderson_extensible_2003}}   &                                & XML                         & Informal                       & \yes                             &                            &                         & \yes                    & \absn                     & \absn                      & \prtl                               & \absn                             & \absn                          & \absn                              & \absn                               \\ %\hline
		\multicolumn{1}{l|}{EPAL \citep{ashley_enterprise_2003}}      &                                & XML                         & Informal                       & \yes                             & Comparison                 &                         & \yes                    & \absn                     & \absn                      & \cmplt                              & \prtl                             & \absn                          & \absn                              & \prtl                               \\ %\hline
		\rowcolor{Gray}
		\multicolumn{1}{l|}{CI \citep{barth_privacy_2006}}            &                                & Formal                      & Formal                         &                                  &                            & \yes                    & \yes                    & \prtl                     & \prtl                      & \prtl                               & \cmplt                            & \prtl                          & \absn                              & \absn                               \\ %\hline
		\multicolumn{1}{l|}{SIMPL \citep{le_metayer_formal_2008}}     & \yes                           & Formal                      & Formal                         &                                  &                            & \yes                    & \yes                    & \prtl                     & \absn                      & \cmplt                              & \cmplt                            & \cmplt                         & \prtl                              & \cmplt                              \\ %\hline
		\rowcolor{Gray}
		\multicolumn{1}{l|}{S4P \citep{becker_s4p_2010}}              & \yes                           & Formal                      & Formal                         &                                  &                            & \yes                    & \yes                    & \prtl                     & \absn                      & \cmplt                              & \cmplt                            & \absn                          & \absn                              & \prtl                               \\ %\hline
		\multicolumn{1}{l|}{Jeeves \citep{yang_language_2012}}        &                                & Formal                      & Formal                         & \yes                             &                            &                         & \yes                    & \prtl                     & \absn                      & \prtl                               & \prtl                             & \absn                          & \prtl                              & \absn                               \\ %\hline
		\rowcolor{Gray}
		\multicolumn{1}{l|}{P2U \citep{iyilade_p2u_2014}}             & \yes                           & XML                         & Informal                       &                                  &                            & \yes                    &                         & \absn                     & \absn                      & \cmplt                              & \cmplt                            & \absn                          & \absn                              & \cmplt                              \\ %\hline
		\multicolumn{1}{l|}{QPDL \citep{van_de_ven_qualitative_2016}} &                                & Formal                      & Formal                         &                                  &                            & \yes                    & \yes                    & \prtl                     & \cmplt                     & \cmplt                              & \cmplt                            & \prtl                          & \cmplt                             & \absn                               \\ %\hline
		\rowcolor{Gray}
		\multicolumn{1}{l|}{RBAC \cite{SCFYrbacm96}}                  &                                & XML                         & Informal                       & \yes                             &                            &                         & \yes                    & \absn                     & \absn                      & \prtl                               & \absn                             & \absn                          & \absn                              & \absn                               \\ %\hline
		\multicolumn{1}{l|}{OSL~\cite{HPBSWplduc07}}                  &                                & Formal                      & Formal                         & \yes                             &                            &                         & \yes                    & \prtl                     & \absn                      & \cmplt                              & \cmplt                            & \absn                          & \cmplt                             & \cmplt                              \\ %\hline
		\rowcolor{Gray}
		\multicolumn{1}{l|}{GeoXACML~\cite{geoaxcml}}                 &                                & XML                         & Informal                       & \yes                             &                            &                         & \yes                    & \absn                     & \cmplt                     & \prtl                               & \prtl                             & \absn                          & \absn                              & \absn                               \\ %\hline
		\multicolumn{1}{l|}{A-PPL~\cite{appl}}                        &                                & XML                         & Informal                       &                                  &                            &                         & \yes                    & \absn                     & \prtl                      & \cmplt                              & \cmplt                            & \prtl                          & \prtl                              & \prtl                               \\ %\hline
		\rowcolor{Gray}
		\multicolumn{1}{l|}{LPL~\cite{LPLgbkb18}}                     & \yes                           & XML                         & Informal                       & \yes                             &                            & \yes                    & \yes                    & \absn                     & \absn                      & \cmplt                              & \cmplt                            & \absn                          & \prtl                              & \cmplt                              \\ %\hline
		\multicolumn{1}{l|}{PrivacyAPIs~\cite{MGLpaactavlpp06}}       &                                & Formal                      & Formal                         &                                  & Analysis                   & \yes                    & \yes                    & \prtl                     & \absn                      & \cmplt                              & \cmplt                            & \cmplt                         & \cmplt                             & \cmplt                              \\ %\hline
		\rowcolor{Gray}
		\multicolumn{1}{l|}{PrivacyLFP~\cite{DGJKDelshgpl10}}         &                                & Formal                      & Formal                         &                                  &                            & \yes                    & \yes                    & \prtl                     & \absn                      & \cmplt                              & \cmplt                            & \cmplt                         & \cmplt                             & \cmplt                              \\ %\hline
		\multicolumn{1}{l|}{PILOT~\cite{pilot_apf_2019}}              & \yes                           & Formal                      & Formal                         &                                  & Analysis                   & \yes                    & \yes                    & \cmplt                    & \cmplt                     & \cmplt                              & \cmplt                            & \prtl                          & \absn                              & \cmplt                              \\ %\hline
		\multicolumn{1}{l|}{}                                          & \multicolumn{5}{c|}{{\scriptsize  Features}}                                                                                                                  & \multicolumn{2}{c|}{{\scriptsize  Audience}}      & \multicolumn{2}{c|}{{\scriptsize  Conditions}}         & \multicolumn{5}{c|}{{\scriptsize  Content}}                                                                                                                                         \\
		\bottomrule % \cline{2-15}    
	\end{tabular}}
	\captionsetup{singlelinecheck=off}
	\caption[foobar]{
		Categorization of privacy languages. The columns indicate:
		\emph{Usability,} whether the language is intended to be understood by DS;
		\emph{Syntax,} whether the syntax of the language defined in XML or a formal language;
		\emph{Enforcement,} whether the language has a formally or informally defined enforcement;
		\emph{Implemented,} whether the language has been implemented;
		\emph{Tools,} the type of available tools for the language;
		\emph{Audience,} whether the language can describe a DS or a DC policy;
		\emph{Conditions,} whether the language supports conditional rules describing when and/or where data may be collected;
		\emph{Content,} whether the language can express the items described in~\Cref{sec:nl_content}.
		We use \cmplt\ to denote that the item is explicitly included in the language;
		\prtl\ to denote that the item is partially supported; and %, \textit{e.g.,} may encoded through conditions or obligations; and
		\absn\ the item is not present in the language and cannot be encoded.
		Some items of the taxonomy are omitted since no solution includes them.
		%
		% \begin{description}
		%   \setlength\itemsep{0.1em}
		% \item[Features] Machine readable privacy policies specific features:
		%   \begin{itemize}
		%     \setlength\itemsep{0.1em}
		%   \item[\textbf{Usability}] Whether the language is \textit{intended} to be understood by DS.
		%   \item[\textbf{Syntax}] Whether the syntax of the language defined in XML or a formal language.
		%   \item[\textbf{Enforcement}] Whether the language has a formally or informally defined enforcement.
		%   \item[\textbf{Implemented}] Whether the language has been implemented.
		%   \item[\textbf{Tools}] This column specifies type of available tools for the language.
		%   \end{itemize}
		% \item[Audience] Whether the language can describe a DS or a DC policy.
		% \item[Conditions] Whether the language supports conditional rules describing when and/or where data may be collected.
		% \item[Content] Whether the language can express the items described in Section \ref{sec:nl_content}.      
		%   %
		%   We use \cmplt\ to denote that the item is explicitly included in the language;
		%   %
		%   \prtl\ to denote that the item is partially supported, \textit{e.g.,} may encoded through conditions or obligations; and
		%   % 
		%   \absn\ the item is not present in the language and cannot be encoded.
		%   %
		%   Note some items of the taxonomy are omitted since no solution includes them.
		%   % \end{itemize}
		% \end{description}
		
%		\vspace*{-.75cm}
	}
	\label{tab:framework_languages}
\end{table}

%%% Local Variables:
%%% mode: latex
%%% TeX-master: "../csur"
%%% End:

\section{Insights}
\label{sec:perspectives}
In this section, we provide several insights that we identified as a
result of our study.\footnote{We refer the reader to
  \Cref{tab:summary} in \Cref{sec:figures} for an overview of the
  juxtaposition of the results combining all facets.}
%
% As we have seen in the previous sections, each facet has a number
% of limitations and benefits.
%
%
We show that each facet is tailored to a specific audience, and
that this is both
\begin{inparaenum}[1)]
\item what makes it beneficial, but also
\item an obstacle to the compliance with all the requirements stated
  in Section~\ref{sec:intro} (\ie, legal validity, understandability
  by all parties, and enforceability through auditable mechanisms).
\end{inparaenum}
In Section~\ref{subsec:intrinsic}, we discuss why a single
facet cannot comply with every requirement.
In Section~\ref{subsec:overcoming_limitations}, we put in perspective
the works which attempt to overcome the limitations of
mono-faceted solutions, and provide guidelines for designing
privacy policies that aim to cover the three facets.
Finally, in Section~\ref{subsec:missing-items}, we discuss the
coverage of the items in our taxonomy by privacy policies in each facet.

\subsection{Limitations of mono-faceted solutions}
%\label{subsec:specificities}
\label{subsec:intrinsic}

A single facet cannot cover all the requirements of privacy policies.
% --- \ie legal validity, understandability by all parties, and enforceability.
%
This is due to the tension between: 
\begin{inparaenum}[i)]
\item the suitability for lawyers,
\item the suitability for lay-users, and
\item the automatic enforcement by machines.
\end{inparaenum}
Concretely, there are details that only have meaning in one facet
and are irrelevant in others.
%
%In the example above, the namespace in the machine-readable policy has no meaning or interest in the other facets.
%
% \Nlpp include details related to compliance with data protection
% regulations, which are unnecessary for the machine-readable facet.
%
Details specific to \nlpp are used by lawyers to check that the policy
complies with privacy protection regulations --- such as the GDPR.
%--- or they refer to functionalities of the system --- \eg logging or cookie management.
%
In general, lay-users may not have the knowledge to fully understand
these details, which makes it less accessible for them.
Likewise, \nlpp do not include low level details related to the
enforcement of the policies by a machine --- those details are often
unnecessary for law enforcement and require lawyers to be familiar
with those technicalities.
%
%\Gpp aim at providing a simplified version of the policy to lay-users.
%
%These policies do not contain details related to the legal aspects of
%the policy, nor aim to be automatically enforced by a machine.
%
\Gpp have the objective of being understood by a general audience, but
this form of privacy policies have no use for lawyers or enforcement
by machines.
\Mrpp aim at being enforced by machines.
They are written in a machine-readable format, and they include all the
necessary details for the underlying system to enforce them.
These details make them difficult to understand by humans, and are,
consequently, unsuitable for lawyers and lay-users.

\paragraph{Illustrative example}
%%% Illustrative example
%
% In order to illustrate the differences in the details that the
% different facets capture, we use an example of privacy policies
% regarding retention time.
%
Consider the
icon~\raisebox{-.35\height}{\includegraphics[height=3.5\fontcharht\font`\B]{conservation_one_year.png}}
from Privacy Tech Icons (see~\Cref{sec:graphical}) that denotes that
data is deleted after 1 year, and this excerpt of Facebook's privacy
policy:
%
% \begin{quote}
  ``[...] when you search for something on Facebook, you can access and
  delete that query from within your search history at any time, but
  \emph{the log of that search is deleted after six months.}''
% \end{quote}
%
Facebook's policy is more precise than the icon: it refers to concrete
data which is produced after a certain user action.
However, these details may not be of prime interest for some lay-users, at least in a first stage.
For instance, they may not know what a log entry or a query are.
Hence, this level of detail may be counter-productive for lay-users.
Yet, this information is required to determine whether Facebook is
processing data according to the policy.
Thus, it cannot be omitted for legal purposes or for users who may be interested in more detailed information.
\lstdefinelanguage{XML}
{
  basicstyle=\small\ttfamily,
  breaklines=true,
  morestring=[s]{"}{"},
  morecomment=[s]{<?}{?>},
  stringstyle=\color{Green},
  identifierstyle=\color{Blue},
  keywordstyle=\color{Cyan},
  morekeywords={xmlns,version,type}% list your attributes here
}
Consider also an excerpt of the policy in APPEL-P3P:
\lstinline[language=XML]{<retention-time days=182 xmlns=".../P3P/retention-time/"/>}.
This policy includes details that are not present in the natural
language policy above:
The format of the policy, XML; the parameter
\lstinline[language=XML]{xmlns}, required so that the computer can
retrieve the set of possible values for the element in the policy (the
XML namespace); and the fact that retention time must be specified in
days.
These details are often irrelevant for law enforcement, and make the
policy difficult to understand for lawyers and lay-users.

\subsection{Multi-faceted privacy policies}
\label{subsec:overcoming_limitations}
% \subsection{Overcoming limitations}
% \label{subsec:overcoming_limitations}
% As argued in Section~\ref{subsec:intrinsic}, a privacy policy expressed in only one facet cannot satisfy requirements for lay-users, lawyers, and auditors.
In many cases, limitations in one facet can be addressed by other facets.
For instance, \nlpp may use \gpp to enhance readability, and \mrpp to be automatically enforced.
%
% Similarly, \mrpp can use \nlpp in order for the latter to be automatically enforced by DC.
%
%Similarly, analysis tools provided with \mrpp can be combined with \gpp to enhance the presentation of risks to DS.
%
In this section, we study \emph{multi-faceted privacy policies}, \ie,
policies that combine any of the three facets studied in this
paper: natural language, graphical, and machine-readable.
%
% To mitigate this issue, we provide guidelines to define policies
% covering the three facets studied in this paper: natural language,
% graphical, and machine-readable.
%
We present existing works on multi-faceted privacy policies, discuss
challenges in defining them, and provide guidelines on addressing
those challenges.

%The current section aims to show that it is possible to combine different facets to overcome limitations.
%In other words: these three facets must be seen as complementary ways to express privacy policies, and they should be used together in order to meet the requirements stated in Section~\ref{sec:intro}.

\subsubsection*{Existing works on multi-faceted privacy policies.}
Several initiatives are already proposing multi-faceted solutions.
Harkous \etal~\cite{harkous_polisis_2018} combine natural language and \gpp, bringing together the accessibility of icons and simple text with the legal value of a natural language privacy policy.
Similarly, \cite{van_den_berg_what_2012,abrams_multilayered_2005}
follow a layered approach to combining graphical and natural language
policies.
Users are first presented with a graphical or simplified version of the policy, which can be refined in several steps all the way to the legal notice.
%
% Policies can be more easily understood thanks to the results of the automatic analysis of \nlpp.
%
SIMPL~\cite{le_metayer_formal_2008}, PILOT~\cite{pilot_apf_2019} and LPL~\cite{LPLgbkb18} combine natural language and \mrpp.
They provide an enforceable policy that helps DS to better understand their choices and provide informed consent --- as required by the GDPR.
%
% \citet{le_metayer_formal_2008} proposed a combination of \nlpp~and \mrpp: a machine-readable formal privacy language, enforceable, but close enough to natural language to be readable.
% %
% PILOT~\cite{pilot_apf_2019} also combines \mrpp~and \nlpp by providing a natural language user interface for DS to input their \mrpp.In
% %
% In LPL~\cite{LPLgbkb18}, it is mandatory to attach a natural language privacy policy to each policy.
%
Similarly, \cite{kelley_nutrition_2009,rossiDaPISOntologyBasedData2019} add graphical representations for P3P policies, resulting in intelligible and enforceable privacy policies.
%
% Kelley \etal~\cite{kelley_nutrition_2009} added a graphical representation on top of P3P --- a machine-readable privacy policy --- resulting in both intelligible and enforceable privacy policies.
% %
% Similarly, Rossi and Palmirani~\cite{rossiDaPISOntologyBasedData2019} based their icon set on an ontology, \ie, they combined a graphical and a machine-readable approach.
%
%Icons are then machine-readable, even though they are not upheld by a formal semantics.
%These examples show that the combination of two facets makes it possible to take advantage of each facet, without loosing benefits.
%
% Other initiatives attempted to dissociate the facets within the same solution. For instance, \citet{van_den_berg_what_2012} implemented a privacy wheel whose spokes present how data is handled. Those spokes have different layers: a first graphical layer aimed at DS, with the possibility to access two other layers of information comprising more details (``legalistic'' information). Another prominent example is the multi-layer approach of \citet{abrams_multilayered_2005}. They proposed a first layer as a short notice of whether data is collected, a second layer as a condensed notice of the data practices in a common graphic format, and finally a third layer, also called full notice, aimed at lawyers. However, these last two initiatives do not consider the machine-readability, and therefore the enforceability.
% of privacy policies.

\begin{figure}[t!]
%\scalebox{.6}{
\centering
\begin{tikzpicture}
\begin{scope}[blend group = hard light]
\fill[red!30!white]   ( 90:1) circle (2);
\fill[green!30!white] (210:1) circle (2);
\fill[blue!30!white]  (330:1) circle (2);
\end{scope}
\node at (90:2)    {Natural Language};
\node at (210:2.2) [align=left]{Machine\\Readable};
\node at (330:2.2) {Graphical};
\node at (0:0)     {None};
\node at (390:1.45) [rotate=315] {\cite{van_den_berg_what_2012,abrams_multilayered_2005,harkous_polisis_2018}};
\node at (270:1.45) [rotate=0]   {\cite{le_metayer_formal_2008,LPLgbkb18,pilot_apf_2019}};
\node at (150:1.45) [rotate=45]  {\cite{kelley_nutrition_2009,rossiDaPISOntologyBasedData2019}};
\end{tikzpicture}
\caption{Works on multi-faceted privacy policies grouped by combination of facets.}
\label{tab:existing-multi-faceted}
%\vspace{-3mm}
\end{figure}
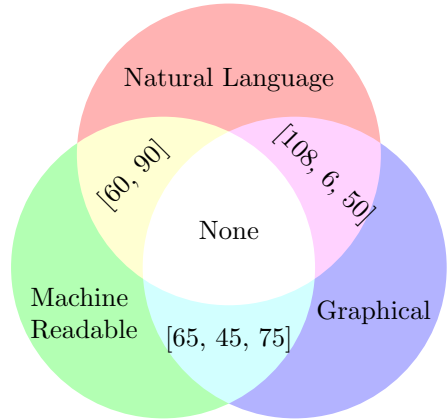

\Cref{tab:existing-multi-faceted} summarizes the landscape of work on
multi-faceted policies.
Note that existing solutions combine at most two types of privacy
policies, and their number is small.
Notably, no existing solution encompasses the requirements for legal
compliance, understandability, and enforceability.

\subsubsection*{Guidelines and challenges}
\label{sec:reco}
Here we set guidelines and discuss challenges in designing
multi-faceted privacy policies, ideally covering all three facets.
We describe two approaches: \emph{unified} and \emph{compound}.
%
% As we have seen, no existing solution encompasses the requirements for
% legal compliance, understandability, and enforceability.
%
% To mitigate this issue, we provide guidelines to define policies
% covering the three facets studied in this paper: natural language,
% graphical, and machine-readable.
%
% We denote these policies \emph{multi-faceted privacy policies}.
%

%% Unified multi-faceted
%%% Guidelines
\paragraph{Unified}
In a unified approach, a core facet is defined and the remaining facets
are generated from the core using a \emph{policy generator}.
Natural language is a suitable candidate as core facet --- it is a legal
requirement and cannot be omitted.
\Mrpp\ could also be considered as core facet as natural language
could automatically be generated from them.
Graphical representations are not suitable for this purpose; they lack
the precision required to generate natural language or \mrpp.
%
%%% Challenges
%
The main challenge of this approach is ensuring the soundness of the
generated policies.
Existing solutions compromise the expressiveness of a facet to
generate policies in other facets.
For example, SIMPL uses constrained natural language so that privacy policies can be enforced by a machine.
As mentioned above, each facet has details that are not addressed by other facets (see~Section~\ref{subsec:intrinsic}).
Thus, another requirement for a policy generator is to include the details specific to the facets generated, although absent in the core facet.
%
% is that the policy generator includes facet
%specific details not present in the core facet when producing a policy
%for another facet.
%
% In order for the the unified approach to be feasible, the policy
% generator must contain the specific details that are necessary for
% each facet: a natural language facet needs legal details, a graphical
% representation must focus on understandability, and a machine-readable
% facet has to include technical details.
%
%As mentioned above, each facet has details that are not addressed by other facets (cf.~Section~\ref{subsec:intrinsic}).
%
%However, these specific details are necessary to each facet: a natural language facet needs legal details, a graphical representation must focus on understandability, and a machine-readable facet has to include technical details.
%
To sum up, the challenge for a unified approach is to ensure that the
facets generated from the core keep what makes them beneficial and
preserve the meaning of the core facet.
%
%For example, SIMPL uses constrained natural language so that privacy
%policies can be enforced by a machine.
%%
%Indeed, the unified approach suffers from scalability problems due to
%the differences in expressiveness of the different facets.
%%
%While it might be easy to manually find a graphical and machine
%readable policy for a natural language policy, automating this process
%for all natural language policies is challenging.

%% Divided multi-faceted
\paragraph{Compound}
%
%%% Guidelines
The compound approach consists in taking mono-faceted policies and
using them together.
For instance, we could use legal text, icons from DaPIS, and
P3P.
Most existing solutions work this way.
Unlike the unified approach, the facets in the compound approach
include the specific details of each facet --- insofar that it combines
existing mono-faceted solutions.
% insofar that it combines existing solutions with their benefits.
%
%%% Challenges
The main challenge here is ensuring consistency between the different facets, \eg, does the machine-readable facet accurately represents
the legal text?
Currently, consistency is manually checked by designers as it
is a very difficult task to automate.
Tool support and strict guidelines may systematize this process and reduce errors.
The machine-readable facet is suitable for full automation, but we
argue that the natural language and graphical facets require expert
supervision (lawyers and designers, respectively).
In a nutshell, the main challenge is to devise methods ensuring
consistency between the different facets.

\begin{figure}[!t]
	\centering
	\begin{tikzpicture}[scale=0.55, every node/.style={scale=1}]
	\matrix(A) [nodes={rectangle,minimum width=1.3cm, minimum height=0.8cm}]
	{
		\node{~~~}; & \multicolumn{5}{c}{\node{~~~~~~~~Graphical Policies};} & \multicolumn{5}{c}{\node{\;\;\;\;\;\;\;\;Machine-Readable Policies};} &  \multicolumn{5}{c}{\node{~~~~~~~~~Required by Legislations};}   \\

		\node{~}; & \node{\cmplt}; & \node{\prtl}; & \node{\absn};
		& \node{\cmplt}; & \node{\prtl}; & \node{\absn};
		& \node{\cmplt}; & \node{\prtl}; & \node{\absn}; \\

		\node{1st party}; & \HeatmapNode{33} & \HeatmapNode{33} & \HeatmapNode{34} 
		& \HeatmapNode{69} & \HeatmapNode{31} & \HeatmapNode{0}
		& \HeatmapNode{80} & \HeatmapNode{20} & \HeatmapNode{0}  \\

		\node{3rd party}; & \HeatmapNode{17} & \HeatmapNode{50} & \HeatmapNode{33} 
		& \HeatmapNode{47} & \HeatmapNode{26} & \HeatmapNode{27}
		& \HeatmapNode{60} & \HeatmapNode{40} & \HeatmapNode{0} \\
		
		\node{Legal basis}; & \HeatmapNode{5} & \HeatmapNode{0} & \HeatmapNode{95} 
		& \HeatmapNode{0} & \HeatmapNode{0} & \HeatmapNode{100}
		& \HeatmapNode{20} & \HeatmapNode{20} & \HeatmapNode{60}  \\

		\node{DS Rights}; & \HeatmapNode{22} & \HeatmapNode{5} & \HeatmapNode{73}
		& \HeatmapNode{13} & \HeatmapNode{17} & \HeatmapNode{70}
		& \HeatmapNode{60} & \HeatmapNode{20} & \HeatmapNode{20}  \\

		\node{Data Retention}; & \HeatmapNode{28} & \HeatmapNode{11} & \HeatmapNode{61} 
		& \HeatmapNode{30} & \HeatmapNode{43} & \HeatmapNode{27}
		& \HeatmapNode{20} & \HeatmapNode{20} & \HeatmapNode{60} \\
		
		\node{Data Security}; & \HeatmapNode{22} & \HeatmapNode{22} & \HeatmapNode{56} 
		& \HeatmapNode{17} & \HeatmapNode{30} & \HeatmapNode{53}
		& \HeatmapNode{40} & \HeatmapNode{40} & \HeatmapNode{20} \\
		
		\node{Policy Change}; & \HeatmapNode{0} & \HeatmapNode{0} & \HeatmapNode{100} 
		& \HeatmapNode{0} & \HeatmapNode{0} & \HeatmapNode{100}
		& \HeatmapNode{0} & \HeatmapNode{20} & \HeatmapNode{80}  \\
	};    
	% \draw[thick] (-.95,2) -- (-.95,-3.6) ;   
	% \draw[thick] (3.54,2) -- (3.54,-3.6) ;     
	\end{tikzpicture}
	\caption{Coverage of taxonomy items by different types of privacy policies, and the privacy legislations in \Cref{tab:content}.}
	\label{fig:item-coverage}
        \vspace{-4mm}
\end{figure}

\subsection{Missing taxonomy items}
\label{subsec:missing-items}
As we saw in the previous sections, not all items of our
taxonomy are covered by existing solutions.
Here we quantify the coverage of each item by the solutions we have studied.
The aim of this section is to shed light on ``forgotten'' items, and hopefully guide future research on these directions.

Figure~\ref{fig:item-coverage} summarizes our results.
Each cell of the heat map shows the percentage of the studied works
(in a given facet) that cover completely (\cmplt), partially (\prtl)
or neither (\absn) an item of the taxonomy.
For instance, the item 1st party is completely covered by 69\% of
policy languages (forth column, first row in
Figure~\ref{fig:item-coverage}).
%
%Thus, it is the best covered item in the taxonomy.
%
% The right part of the table put the coverage in perspective of the legal requirements presented in Table~\ref{tab:content}.

The 1st and 3rd party taxonomy items have the best coverage in our
study.
Probably due to the fact that they express the most relevant
information regarding data collection and processing for DS.
These items are followed (in terms of coverage) by Data Retention and
Data Security, which are absent from almost 50\% of the languages
studied --- except for Data Retention in machine-readable privacy
policies.
DS Rights is absent in around 70\% of the graphical and
machine-readable policy languages.
Most likely, because they refer to information difficult to express
graphically, and they are outside the realm of what
machine-readable language are designed for.
%
% Nevertheless, work on expressing DS Rights graphically would be of
% great use for DS.
% 
Finally, legal basis and policy change are absent from all the studied work.
Possibly due to their absence in most legislations; 60\% and 80\%
respectively.

\section{Final discussion}
\label{sec:conclusion}

\paragraph{Related work}
This work is not the first to propose an overview of the different
manners to express privacy policies.
Schaub~\etal~\cite{schaub_design_2015} present the requirements and
best practices for presenting privacy \emph{notices}.
Their work focuses on providing users impactful notices.
In other words, they study how well they understand the messages
conveyed by the privacy notices.
%
% In our work, we connect this dimension with its legal validity, \ie,
% what parts of \nlpp\ are covered by privacy notices.
%
Cranor~\cite{cranor_necessary_2012} describes the notice and choice mechanisms, what P3P attempted to do to palliate issues raised by these principles, and why it failed in doing so.
These solutions focus on the design of privacy policies to enhance
usability for lay-users.
In our work, we focus on connections of the graphical and
machine-readable privacy policies with legal requirements (\ie,
\nlpp).
Although both of these articles consider the machine-readibility of privacy
notices, we highlight benefits that they did not consider, such as the
possibility it offers for enforcement, auditing or automatic consent
management.

%
% They mostly review work focused on presenting information to
% lay-users, but do not address how this dimension is distinct from the
% legal validity of privacy policies (notably in natural language), nor
% the different possibilities offered by \mrpp.
%
% However, they provide useful insights, for instance when they
% recommend standardized approaches, or a combination between graphical
% and machine-readable dimensions (although not in these terms).
%
To the best of our knowledge, this work is the first to systematically
study privacy policies based on the different means of expression or
``facets'', and how these facets must be combined to provide legally
valid, usable and enforceable privacy policies.

\paragraph{Conclusion}
In this paper, we have studied the different ways to express privacy
policies: in natural language, with graphical representations, and
using machine-readable means.
We have categorized the existing work in each facet according to
a taxonomy of privacy policies, as well as their specific features.
Additionally, we have studied the benefits and limitations of each
facet, and we have shown that the limitations of one facet
can be addressed by the benefits of the other facets.
We have studied the combination of different facets, which overcomes limitations by bringing together the benefits of each facet, and provided guidelines to design multi-faceted privacy policies.
%
%We have proposed guidelines to express multi-faceted privacy policies, which overcomes the limitations of each facet by bringing together their benefits.
%
%Finally, 
We have made explicit the degree of coverage of the items in our taxonomy by the surveyed work, thus shedding light on future research directions on the design of mono- and multi-faceted privacy policies.
We envision this work as an effort to facilitate and boost collaborative work between the legal domain, design, and computer science, and to provide a big picture of how transparency can be ensured through privacy policies.

\subsubsection*{Acknowledgments} This work has been partially funded by the CHIST-ERA project UPRISE-IoT (User-centric Privacy and Security in the IoT) and the ANR project CISC (Certification of IoT Secure Compilation).
We address special thanks to Daniel Le M\'etayer for his insights and comments on the early stages of this article.

\newpage

\bibliography{PhD,PhD2}
%\bibliography{}
\bibliographystyle{plainnat}

\appendix
\onecolumn
\section{Appendix}
\label{sec:figures}

\Cref{tab:summary} provides a condensed overview of the content, benefits, limitations and tools that we have studied for each facet.
The content columns are a coarsed grained presentation of the results in \Cref{fig:item-coverage}.
The three last columns list the benefits, limitations and tools of each facet.

% \newcolumntype{L}[1]{>{\raggedright\let\newline\\\arraybackslash\hspace{0pt}}m{#1}}
% \newcolumntype{C}[1]{>{\centering\let\newline\\\arraybackslash\hspace{0pt}}m{#1}}
% \newcolumntype{R}[1]{>{\raggedleft\let\newline\\\arraybackslash\hspace{0pt}}m{#1}}

\begin{table}[h!]
\scalebox{.7}{
  \begin{tabular}{l|ccccccccl|l|l|l|}
  \multicolumn{1}{c}{}                 & \multicolumn{1}{R{45}{1em}}{1st Party}  & \multicolumn{1}{R{45}{1em}}{3rd Party}  & \multicolumn{1}{R{45}{1em}}{Legal basis} & \multicolumn{1}{R{45}{1em}}{DS rights} & \multicolumn{1}{R{45}{1em}}{Data retention} & \multicolumn{1}{R{45}{1em}}{Data security} & \multicolumn{1}{R{45}{1em}}{Policy change} & \multicolumn{1}{R{45}{1em}}{Others} & \multicolumn{1}{c}{~~} & \multicolumn{1}{c}{}         &    \multicolumn{1}{c}{}   & \multicolumn{1}{c}{}   \\
  \midrule
  \rowcolor{Gray}
Natural Language          & \yes & \yes & \yes        & \yes      & \yes           & \yes          & \yes          & \yes   &~~& - Legal value                & - Ambiguity                     & - Templates                               \\
\rowcolor{Gray}           &      &      &             &           &                &               &               &        &~~&                              & - Understandability             & - Generators                              \\
\rowcolor{Gray}           &      &      &             &           &                &               &               &        &~~&                              & - Enforceability                & - Retrievers                              \\
\rowcolor{Gray}           &      &      &             &           &                &               &               &        &~~&                              &                                 & - Analysis tools                          \\
\rowcolor{Gray}           &      &      &             &           &                &               &               &        &~~&                              &                                 &                                           \\
Graphical                 & \yes & \yes & \yes        & \yes      & \yes           &               &               &        &~~& - Understandability          & - Ambiguity                     & - DS Notification                         \\
                          &      &      &             &           &                &               &               &        &~~&                              & - Incompleteness                & - Visual comparison                       \\
                          &      &      &             &           &                &               &               &        &~~&                              &                                 &                                           \\
\rowcolor{Gray}
Machine Readable          & \yes & \yes &             & \yes      & \yes           &               &               &        &~~& - Enforcement                & - Understandability             & - Enforcement engines                     \\
\rowcolor{Gray}           &      &      &             &           &                &               &               &        &~~& - Auditability               & - Lack of adoption              & - Formal Semantics                        \\
\rowcolor{Gray}           &      &      &             &           &                &               &               &        &~~& - Correctness                &                                 & - Policy comparison                       \\
\rowcolor{Gray}           &      &      &             &           &                &               &               &        &~~& - Automation                 &                                 & - Analysis tools                          \\
\rowcolor{Gray}           &      &      &             &           &                &               &               &        &~~&                              &                                 &                                           \\
                          &      &      &             &           &                &               &               &        &~~&                              &                                 &                                           \\
        \multicolumn{1}{c|}{~} & \multicolumn{8}{c}{Content}                                                                     &~~& \multicolumn{1}{c|}{Benefits} & \multicolumn{1}{c|}{Limitations} & \multicolumn{1}{c|}{Tools}                 \\    
\bottomrule
\end{tabular}}
\caption{Summary of each facet's content, benefits, limitations and tools. The \yes symbol indicates that at least one work on the corresponding facet captures the taxonomy item.}
\label{tab:summary}
\end{table}

\definecolor{gray}{rgb}{0.4,0.4,0.4}
\definecolor{darkblue}{rgb}{0.0,0.0,0.6}
\definecolor{cyan}{rgb}{0.0,0.6,0.6}

\lstset{
	basicstyle=\scriptsize\linespread{0.4}
}

\lstdefinelanguage{XML}
{
	morestring=[b]",
	morestring=[s]{>}{<},
	morecomment=[s]{<?}{?>},
	stringstyle=\color{black},
	identifierstyle=\color{darkblue},
	keywordstyle=\color{cyan},
	morekeywords={xmlns,version,type}% list your attributes here
}

% \begin{marginfigure}[-25em]

% \end{marginfigure}
% \begin{figure}
% 	% \centering
% 	\definecolor{lemonchiffon}{rgb}{0.94, 0.88, 0.19}
% 	\sethlcolor{lemonchiffon}
% 	\begin{tikzpicture}
% 	% 
% 	\node[text width=7cm] (nlp) {\footnotesize ``\hl{We share
% 			information globally, both internally within the Facebook
% 			Companies and externally with our partners and with those you
% 			connect and share with around the world in accordance with
% 			this Policy.}  Information controlled by Facebook Ireland will
% 		be transferred or transmitted to, or stored and processed in,
% 		the United States or other countries outside where you live for
% 		the purposes as described in this Policy.[...]''};
% 	% 
% 	\node[above=of nlp, yshift=-7mm, xshift=-25mm, fill=lemonchiffon] (gp) {\includegraphics[scale=.3]{Figures/raskin_Reuse_thirdparty.png}};
% 	\node[above=of nlp, yshift=-7mm, xshift=15mm, fill=lemonchiffon, align=left, font=\tiny] (mr1) {
% 		\scriptsize$(\mathit{Facebook},\ldots,$\\
% 		\scriptsize$~~\{~(\mathit{Internal\_Companies},\ldots),$\\
% 		\scriptsize$~~~~~(\mathit{External\_Parties},\ldots)\})$
% 	};
% 	\node[above=of nlp, xshift=-2mm, yshift=-14mm] (aux1) {};
% 	\node[above=of nlp, xshift=32mm, yshift=-18mm] (aux2) {};
% 	\draw[<-, color=lemonchiffon, ultra thick] (gp) -- (aux1);
% 	\draw[<-, color=lemonchiffon, ultra thick] (mr1) -- (aux1);
% 	% \draw[-, color=lemonchiffon, ultra thick] (mr2) -- (aux2);
% 	\end{tikzpicture}
% 	\caption{Multifaceted Privacy Policy Overview.}
% 	\label{fig:multifaceted-policy}
% \end{figure}

\end{document}